   \documentclass[twocolumn,showpacs,preprintnumbers,amsmath,amssymb]{revtex4-2}
   \usepackage{graphicx,xcolor}    
\usepackage{epstopdf,epsfig}
	\usepackage{bm}
   \usepackage{dcolumn}
   \usepackage{natbib}
	\usepackage{physics}

\begin{document}
   	
   	\title{Excitonic Tunneling in the AB-bilayer Graphene Josephson Junctions}
  
   	\author{V. Apinyan\footnote{Corresponding author. Tel.:  +48 71 3954 284; E-mail address: v.apinyan@int.pan.wroc.pl.}, T. K. Kope\'{c}}
   	\affiliation{Institute for Low Temperature and Structure Research, Polish Academy of Sciences\\
   		PO. Box 1410, 50-950 Wroc\l{}aw 2, Poland \\}
   	
   	\date{\today}

\begin{abstract}
%
We have considered the AB-stacked bilayer graphene Josephson junction. The bilayers are supposed to be in the charge equilibrium states and at the half-filling in each of the electronic layers of the construction and at each value of the external gate. By considering the interacting bilayers in both sides of the junction and by taking into account both intralayer and interlayer Coulomb interaction effects, we have calculated the normal and excitonic tunnel currents through the junction. The electronic band renormalizations have been taken into account, due to the excitonic pairing effects and condensation in the BLGs. The exact four-band energy dispersions, including the excitonic renormalizations, have been used for the bilayers without any low-energy approximation. We show the degeneracy of the ground state at the zero applied voltage and for different phases of the coherent condensates by showing a dc Josephson current through the junction. The normal and excitonic tunneling currents have been calculated for different gate voltages and for different values of the interaction parameters. The role of the charge neutrality point has been discussed in details.      
\end{abstract}

   	\maketitle

 \renewcommand\thesection{\arabic{section}}
   	
\section{\label{sec:Section_1} Introduction}
%
The existence of the electron-hole bound states for a semimetal with overlapping bands has been postulated long years ago by Keldysh and Kopaev \cite{cite_1}, and the prediction about the superfluidity has been given for a
condensed excitonic state and ulteriorly, it  has been the subject of the intense theoretical studies \cite{cite_2, cite_3,cite_4,cite_5,cite_6,cite_7,cite_8,cite_9, cite_10, cite_11}. Experimentally, the strong evidence of an excitonic insulator (EI) and excitonic Bose-Einstein condensate (BEC) ground states has been shown only in the quantum Hall regime (in a large magnetic field) and under the high pressure in a series of the experimental works on the rare-earth chalcogogenide compounds, transition metal dichalcogenides and tantalum chalcogenides \cite{cite_12}. The exciton condensation was experimentally observed also in quantum Hall bilayers \cite{cite_13,cite_14}, in the systems of magnons \cite{cite_15} and cavity exciton polaritons \cite{cite_16,cite_17}.
Recently, other solid state systems were proposed as possible candidates for the
achievement of the BEC of excitons. It concerns the quantum well heterostructures with the excitons trapped in the cavities of the potential wells \cite{cite_18,cite_19, cite_20, cite_21, cite_22, cite_23, cite_24, cite_25, cite_26,cite_27, cite_28, cite_29} (the structure utilized in these works were double-layer GaAs/AlGaAs or InAs/GaSb quantum-wells with an electric field applied perpendicularly to the structure), the double layer heterostructures and the bilayers \cite{cite_30, cite_31, cite_32}   

The excitonic gap formation and the condensation has been examined also in the bilayer graphene structures \cite{cite_33, cite_34, cite_35, cite_36,cite_37,cite_38}. Namely, the bilayer graphene is very promising for the optoelectronic applications due to its unique gate-controllable band structure properties \cite{cite_39}. The imposition of external electrical field can tune the bilayer graphene from the semimetal to the semiconducting state. Nevertheless, the excitonic condensation in the bilayer graphene structures remains
controversial in the modern solid state physics because of the complicated nature of the single-particle correlations in these
systems \cite{cite_33, cite_34, cite_35, cite_36, cite_37, cite_38}. It has been shown recently \cite{cite_40} that  the critical temperature, which describes the transition from the condensate state to the normal state in graphene double layer structure, can be very high due to the extremely small effective mass of excitons. The coherence in exciton BEC condensates survives at the very high temperatures. An analogue conclusion has been drawn in Ref.\onlinecite{cite_38}, concerning the bilayer graphene, where the condensate evolution has been analysed as a function of the interlayer Coulomb interaction parameter in the BLG. 

Recently, the excitonic condensation has been realized experimentally in the double bilayer graphene heterostructure in the strong quantum Hall regime and by a combination of Coulomb drag and current counterflow measurements \cite{cite_41}. They have also found the evidence of strong interlayer coupling between the graphene layers thanks to the quantized Hall ``drag plateau''. The zero-valued longitudinal resistance measured there confirms the dissipationless (friction-free) nature of the electron-hole condensate state.
A quite simple experimental way to observe the excitonic condensate states in the bilayer structures is related to the possibility of engineering of a spatially confined excitonic condensates in the potential traps, and the investigation of the Josephson tunneling 
effects for excitons \cite{cite_42}, related to the tunnelling between two trapped Bose condensates \cite{cite_43} that possess a macroscopic phase coherence. The excitonic Josephson tunnelling effects and thermal transport properties in the electron-hole type double layer graphene junctions, separated by a dielectric layer, have been recently considered in Refs.\onlinecite{cite_44, cite_45}. 

In the present paper we study the excitonic tunnelling effects in the tunnel junction based on the AB-stacked bilayer graphene structures. We suppose the presence of the macroscopic phase coherence regime with the well defined condensates phases and amplitudes and we consider only the local on-site interlayer excitonic pairing in each side of the junction. Supposing the electronic bilayers, without the initial optical pumping mechanism, we study the normal and excitonic tunnelling currents through the BLG/I/BLG junction for different values of the external gate voltage applied to the heterostructure. We will assume the half-filling regime in each layer of the BLG structures, even in the presence of the applied gate potential, thus by supposing that not considerable changes of electron density occurs during the adiabatic switching of the external potential. We will show how a finite difference between the phases of the coherent excitonic condensates in the BLG subsystems leads to the excitonic Josephson dc current through the tunnel junction at the zero external voltage. We show also that any finite voltage leads to the ac Josephson current irrespective of the phases of coherent condensates in both sides of the junction. We study the amplitude of the zero voltage Josephson dc current as a function of the interlayer Coulomb interaction parameters in the subsystems. The symmetric and asymmetric interaction cases have been considered straightforwardly. Also, we calculate the normal quasiparticle tunneling current and we show that normal tunneling in the BLG/I/BLG heterostructure is an interaction-protected process and the threshold frequency of the normal tunneling current strongly depends on the values of the Coulomb interaction parameters in the BLGs. We analyse the role of the charge neutrality point (CNP) on the behavior of the normal and excitonic tunneling currents.      
%
\section{\label{sec:Section_2} The bilayer graphene Josephson junction}
%
\subsection{\label{sec:Section_2_1} Description of the Hubbard interactions}
\begin{figure}
		\begin{center}
			\includegraphics[width=500px,height=100px]{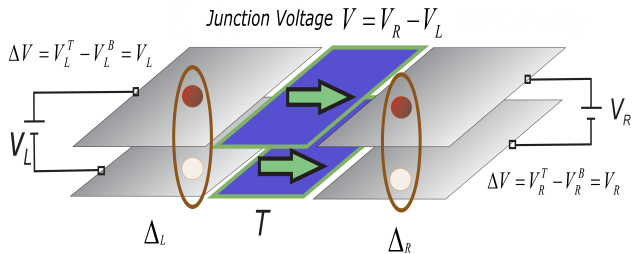}
			\caption{\label{fig:Fig_1}(Color online) The structure of the BLG/I/BLG tunnel junction. The applied gate potentials and the potential drop (between the layers) are shown in the left ($V_L$) and right ($V_{R}$) side of the junction.}.
		\end{center}
	\end{figure}

%
We introduce here our model consisting of two Bernal stacking bilayer graphene (BLG) structures separated by a very thick dielectric layer (we suppose that the thickness of the insulating layer is such that we can neglect the quasiparticle scattering and recombination processes in the layer). For the convenience, we will denote by $a, b$ and $\tilde{a}, \tilde{b}$ (and their conjugates $a^{\dag}, b^{\dag}$ and $\tilde{a}^{\dag}, \tilde{b}^{\dag}$) the annihilation (creation) fermionic operators corresponding to different sublattice sites $A$, $B$ in the bottom, and $\tilde{A}$, $\tilde{B}$ in the top layer of the left-BLG. Similarly, we denote by $c, d$ and $\tilde{c}, \tilde{d}$ (and their conjugates $c^{\dag}, d^{\dag}$ and $\tilde{c}^{\dag}, \tilde{d}^{\dag}$) the annihilation (creation) fermionic operators corresponding to different sublattice sites in the bottom, and top layer of the right-BLG.  
In Fig.~\ref{fig:Fig_1}, we have presented the schematic setup of our BLG/I/BLG junction (here \textit{I} represents the dielectric layer between the BLGs). We assume here that the layers, which have lattice sites $A$ and $B$ (bottom layer) are biased -$V_{\ell}/2$ (with $\ell=L,R$), and the layers with the lattice sites $\tilde{A}$ and $\tilde{B}$ (top layers) $V_{\ell}/2$, so that the potential difference between the two layers is $V_{\ell}$. For the first treatment of such a junction, we suppose the half-filling condition satisfied in both BLG systems, i.e., we suppose that $\langle n^{\ell'}_{\ell} \rangle=1$, where the $n^{\ell'}_{\ell}$ is the total particle number operator in each layer with $\ell'=1,2$ of each BLG with $\ell= L,R$. Note, also that we attach the number $\ell'=1$ to the bottom layers and $\ell'=2$ to the top layers in the heterostructure.   

The non-interacting tight-binding part of the total junction-Hamiltonian could be written in the usual form
	\begin{eqnarray}
	{H}_{0}=\sum_{\ell=L,R}\hat{H}_{\ell 0}
		\label{Equation_1}
	\end{eqnarray}
	with ${H}_{{\ell}0}$, given by
	 \begin{eqnarray}
	 {H}_{{\ell}0}=&&-\gamma_0\sum_{\left\langle {\bf{r}}{\bf{r}}'\right\rangle}\sum_{X_{\ell},Y_{\ell}}\sum_{\sigma}\left(X^{\dag}_{\ell\sigma}({\bf{r}})Y_{\ell\sigma}({\bf{r}}')+h.c.\right)
	 \nonumber\\
	 &&-\gamma_1\sum_{{\bf{r}},\sigma}\left(P^{\dag}_{\ell\sigma}({\bf{r}})Q_{\ell\sigma}({\bf{r}})+h.c.\right)
	 \nonumber\\
	 &&-\sum_{\ell'}\sum_{{\bf{r}},\sigma}\mu^{\ell'}_{\ell}n^{\ell'}_{\ell,\sigma}({\bf{r}}),
	 \label{Equation_2}
	 \end{eqnarray}
	 where the fermionic operators $X_{\ell}$ and $Y_{\ell}$ refer to different sublattice fermions, i.e., for $\ell=L$ we have $X_{\ell}=a, \tilde{a}$, and $Y_{\ell}=b,\tilde{b}$, while for $\ell=R$ we have $X_{\ell}=c, \tilde{c}$ and $Y_{\ell}=d,\tilde{d}$. The intralayer hopping parameter $\gamma_0$ is supposed to be the same for both layers and for both BLGs. The fermionic operators $P^{\dag}_{\ell}$ and $Q^{\dag}_{\ell}$ in Eq.(\ref{Equation_2}) are defined such that for $\ell=L$ we have $P_{\ell}=b$ and $Q_{\ell}=\tilde{a}$, whereas for $\ell=R$ we defined $P_{\ell}=d$ and $Q_{\ell}=\tilde{c}$. The parameter $\gamma_1$ denotes the interlayer hopping amplitude between the layers of the BLGs, and it is supposed also the same for both sides of the junction. The variable $\sigma$, in all terms in Eq.(\ref{Equation_2}), describes the fermionic spin variable, which takes two values $\sigma=\uparrow, \downarrow$. The last term in Eq.(\ref{Equation_2}) subjects the chemical potential terms with the chemical potentials coupled to the coupled charge density operators in each layer of separated BLGs. Initially, we suppose that the BLG/I/BLG system is in the Grand canonical equilibrium state with the equal chemical potentials in both layers of each BLG structure. Thus, we have $\mu^{\ell'=1}_{\ell}=\mu^{\ell'=2}_{\ell}$ for $\ell=L,R$. The charge density operator $n^{\ell'}_{\ell,\sigma}({\bf{r}})$, as it was discussed above, describes the total electron densities with the spin $\sigma$ in the given layer $\ell'=1,2$ of a given BLG with $\ell=L,R$. For example, for the layer $\ell'=1$ in the BLG with $\ell=L$, we have $n^{1}_{L,\sigma}({\bf{r}})=a^{\dag}_{\sigma}({\bf{r}})a_{\sigma}({\bf{r}})+b^{\dag}_{\sigma}({\bf{r}})b_{\sigma}({\bf{r}})$. It is important to notice here that we consider purely electronic graphene layers in both sides of the heterojunction and we do not suppose any initial optical pumping in the BLGs.  
	 %
	%
	The fermionic operators in the separate BLG systems satisfy the following anticommutation rules
		\begin{eqnarray}
		\left[X^{l'}_{l}({\bf{r}}),Y^{m'}_{m}({\bf{r}}')\right]_{+}&=&X^{l'}_{l}({\bf{r}})Y^{m'}_{m}({\bf{r}}')+Y^{m'}_{m}({\bf{r}}')X^{l'}_{l}({\bf{r}})
		\nonumber\\
		&&=\delta_{ll'}\delta_{mm'}\delta(X-Y)\delta({\bf{r}}-{\bf{r}}').
			\nonumber\\
		\label{Equation_3}
		\end{eqnarray}
	The top indices in Eq.(\ref{Equation_3}) indicates the layers in the BLGs, and the bottom subscripts were introduced for the left- and right-BLGs, as in Eq.(\ref{Equation_2}). Thus, we suppose the nonvanishing anticommutators only between the fermions on the same type of sublattices in the layers. The sign $+$ near the parenthesis $[...]_{+}$, in the left-hand side in Eq.(\ref{Equation_3}), indicates the anticommutation rule.   	
	
	We introduce here also the interaction terms by defining the generalized bilayer Hubbard model in each side of the junction. Namely, we have for the interaction part
		\begin{eqnarray}
		{H}_{\rm int}=\sum_{\ell=L,R} {H}_{i\ell},
		\label{Equation_4}
		\end{eqnarray}
		where ${H}_{i\ell}$ are the interaction Hamiltonians, for both sides of the construction
		\begin{eqnarray}
		{H}_{i\ell}=U_{\ell}\sum_{\ell'}\sum_{{\bf{r}},\eta}\left[\left(n^{\ell'}_{\ell\eta\uparrow}-\frac{1}{2}\right)\left(n^{\ell'}_{\ell\eta\downarrow}-\frac{1}{2}\right)-\frac{1}{4}\right]
		\nonumber\\
		+W_{\ell}\sum_{{\bf{r}}\sigma\sigma'}\left[\left(n^{\ell'=1}_{P_{\ell}\sigma}-\frac{1}{2}\right)\left(n^{\ell'=2}_{Q_{\ell}\sigma'}-\frac{1}{2}\right)-\frac{1}{4}\right],
		\label{Equation_5}
		\end{eqnarray}
	 where $n^{\ell'=1}_{P_{\ell}\sigma}$ and $n^{\ell'=2}_{Q_{\ell}\sigma}$ are the electron densities in the layers 1 and 2 in the BLGs for the $P_{\ell}$ and $Q_{\ell}$-type fermions, introduced above. Then we will use the interaction representation for the fermions \cite{cite_46}, in which the time dependence of the fermionic operators is given by the unperturbed Hamiltonian ${H}_{L 0}(V)+{H}_{R 0}$, where ${H}_{L 0}(V)$ is the Hamiltonian of the left side of the junction in the presence of the external gate voltage applied to the BLG/I/BLG. We assume that the gate voltage $V(t)$ drops across the barrier and, in general, the Fermi levels in the left and right-BLG structures will relatively shift by an amount proportional to the potential drop across the junction, i.e., $\bar{\mu}_{L}-\bar{\mu}_{R}\sim-eV(t)$. For the reasons that will be clear in the following sections, we have denoted by $\bar{\mu}_{l}, l=L,R$ the exact Fermi levels in different sides of the tunnel junction. We suppose here the half-filling conditions for the total electron densities in each layer of the separate BLG structure and we assume that the influence of the external gate voltage on the charge densities in the layers is infinitesimally small, and the excitonic gap parameter will not get modified by the external perturbation, i.e., $\delta{V}\nrightarrow \delta{n^{\ell'}_{\ell}}\nrightarrow \delta{\Delta}$, and consequently $\delta{n^{\ell'}_{\ell}}=0$, $\delta{\Delta}=0$. Such a nonperturbative effect on the electron densities and on the excitonic gap parameter permits to include properly the effect of the applied gate voltage on the excitonic properties in the system, and do not affects the excitonic condensate state in the system. For a more sophisticated case, that evolve the variation of the excitonic condensates states in the junction, one should include the influence of the gate voltage on the excitonic gap parameter and the hal-filling assumption will be failed in this case. Such a treatment is out of scope of the present paper. 
	 %
	 %
	 \subsection{\label{sec:Section_2_2} The tunneling Hamiltonian}
	 %
	 The time evolution of the eigenstates is determined by the perturbation term, given via the quasiparticle tunneling Hamiltonian ${H}_{T}$. In the following, we will transform the total Hamiltonian of the system ${H}={H}_{0}+{H}_{\rm int}+H_{T}$ into the Fourier space by introducing the Nambu fermionic spinors $\psi^{L}_{{\bf{k}},\sigma}(t)$ and $\psi^{R}_{{\bf{k}},\sigma}(t)$ in the left and right sides of the junction. We have 
	\begin{eqnarray}
	\psi^{L}_{{\bf{k}},\sigma}(t)=\left[a_{{\bf{k}},\sigma}(t),b_{{\bf{k}},\sigma}(t),\tilde{a}_{{\bf{k}},\sigma}(t),\tilde{b}_{{\bf{k}},\sigma}(t)\right]^{T},
	\nonumber\\
	\psi^{R}_{{\bf{k}},\sigma}(t)=\left[c_{{\bf{k}},\sigma}(t),d_{{\bf{k}},\sigma}(t),\tilde{c}_{{\bf{k}},\sigma}(t),\tilde{d}_{{\bf{k}},\sigma}(t)\right]^{T}.
\label{Equation_6}
\end{eqnarray}
	 The tunnelling matrix Hamiltonian ${H}_{T}$ will include all possible quasiparticle tunnellings between the BLGs in the junction. Namely, it accounts all sublattice fermions in the hexagonal layers that participate to the tunnelling process. Due to the AB stacking order in the BLGs and the form of the interlayer hopping and interlayer interaction terms in Eqs.(\ref{Equation_2}) and (\ref{Equation_4}), some of the sublattice fermions will not participate to the total excitonic tunnelling (for example the fermions on the lattice sites $a,\tilde{b}$ and $c, \tilde{d}$), and they will contribute only to the single-particle normal tunnelling in the junction. Therefore, for a more general case, which includes the tunnellings from all possible fermionic flavours, we will write ${H}_{T}$ as
	 	\begin{eqnarray}
	{H}_{T}=\sum_{{\bf{k}},{\bf{p}}}\sum_{\sigma}t_{{\bf{k}},{\bf{p}}}\left(\tilde{a}^{\dag}_{{\bf{k}},\sigma}(t)\tilde{c}_{{\bf{p}},\sigma}(t)+{b}^{\dag}_{{\bf{k}},\sigma}(t){d}_{{\bf{p}},\sigma}(t)+\right.
	\nonumber\\
	\left.\tilde{b}^{\dag}_{{\bf{k}},\sigma}(t)\tilde{d}_{{\bf{p}},\sigma}(t)+{a}^{\dag}_{{\bf{k}},\sigma}(t){c}_{{\bf{p}},\sigma}(t)\right),
	\label{Equation_7}
	\end{eqnarray}
			where $t_{{\bf{k}},{\bf{p}}}$ is a matrix element, which describes the transition probability for an electron from the state ${\bf{k}}$ (in the left side of the barrier) to a state ${\bf{p}}$ (in the right side of it). The total tunnelling current through the BLG/I/BLG junction will be expressed as
	\begin{eqnarray}
		I(V,T,t)=-e\langle \dot{N}^{\ell=2}_{R}(t)\rangle,
	 \label{Equation_8}
	 \end{eqnarray}
	 where $\langle \dot{N}^{\ell=2}_{R}(t)\rangle$ is the expectation value of the rate of change of the electron number operator ${N}^{\ell=2}_{R}$ in the top right layer of the heterojunction, i.e.,
	 \begin{eqnarray}
	  {N}^{\ell=2}_{R}(t)=\sum_{{\bf{k}},\sigma}\left(\tilde{c}^{\dag}_{{\bf{k}},\sigma}(t)\tilde{c}_{{\bf{k}},\sigma}(t)+\tilde{d}^{\dag}_{{\bf{k}},\sigma}(t)\tilde{d}_{{\bf{k}},\sigma}(t)\right).
	   \label{Equation_9}
	  \end{eqnarray}
	  The expectation value $\langle \dot{N}_{R}\rangle$ is given by $\langle \dot{N}_{R}\rangle=\Tr(e^{-\beta{H}}\dot{N}_{R})/\Tr(e^{-\beta{H}})$, where $\beta=1/k_{B}T$, and $H$ is the total Hamiltonian of the system $H$. The Heisenberg equation of motion for ${N}^{\ell=2}_{R}$ is 
	 	\begin{eqnarray}
i\hbar\frac{d{N}^{\ell=2}_{R}(t)}{dt}=\left[{N}^{\ell=2}_{R}(t),H(t)\right]_{-}=\left[{N}^{\ell=2}_{R}(t),H_{T}(t)\right]_{-}.
\nonumber\\
	 \label{Equation_10}
	 \end{eqnarray}
	 The notation $[\ldots]_{-}$ refers to the usual Bose-type commutation rule. A very simple calculation shows that
	  	\begin{eqnarray}
	  I(V,T,t)=\frac{2e}{\hbar}\Im\left[\sum_{{\bf{k}},{\bf{p}}}\sum_{\sigma}t_{{\bf{k}},{\bf{p}}}\left\langle\tilde{a}^{\dag}_{{\bf{k}},\sigma}(t)\tilde{c}_{{\bf{p}},\sigma}(t)+\right.\right.
	  \nonumber\\	  	
	  \left.\left.+\tilde{b}^{\dag}_{{\bf{k}},\sigma}(t)\tilde{d}_{{\bf{p}},\sigma}(t)\right\rangle\right].
	  \label{Equation_11}
	  \end{eqnarray}
	 Furthermore, we apply the Ryckayzen \cite{cite_47} transformation for the creation and destruction operators in the left-BLG system. For the top-left and bottom-left layer's electrons, this transformation implies that
	 	\begin{eqnarray}
	 	X^{\ell=2}_{{\bf{k}},\sigma}(t) \rightarrow 	X^{\ell=2}_{{\bf{k}},\sigma}(t)e^{i\Delta{\Phi}_{\rm Upper}(t)},
	 	\nonumber\\
	 	X^{\ell=1}_{{\bf{k}},\sigma}(t) \rightarrow 	X^{\ell=1}_{{\bf{k}},\sigma}(t)e^{i\Delta{\Phi}_{\rm Lower}(t)},
	 	  \label{Equation_12}
	 	\end{eqnarray}
	    where $X^{\ell=2}_{{\bf{k}},\sigma}(t)=\tilde{a}_{{\bf{k}},\sigma}(t), \tilde{b}_{{\bf{k}},\sigma}(t)$ in the top layer, and $X^{\ell=1}_{{\bf{k}},\sigma}(t)={a}_{{\bf{k}},\sigma}(t), {b}_{{\bf{k}},\sigma}(t)$ in the bottom layer. We can suppose that the phase differences across the junction, between the upper and lower layers in the BLGs, evolve with voltage according to the relations known in the usual Josephson junctions theory \cite{cite_48}, and a good reason for this is related to the fact that the total ground state energy of the separate non-interacting graphene layer equals exactly twice of the chemical potential in the layer $E_{\ell}=2|{\mu}_{\ell}|$ with $\ell=L,R$ \cite{cite_49}, which is the case of the usual superconductors according to the Gorkov's derivations of the energy spectrum of superconductors in the microscopic theory \cite{cite_50}. Thus, we have
	\begin{eqnarray}
\Delta{\Phi}_{\rm Upper}(t)=\frac{2e}{\hbar}\int^{t}_{0}dt'\left(\frac{V_{L}(t)-V_{R}(t)}{2}\right)-\frac{{\Delta\varphi}_{0}}{2},
\nonumber\\
\Delta{\Phi}_{\rm Lower}(t)=-\frac{2e}{\hbar}\int^{t}_{0}dt'\left(\frac{V_{L}(t)-V_{R}(t)}{2}\right)+\frac{{\Delta\varphi}_{0}}{2}.
\nonumber\\
 \label{Equation_13}
\end{eqnarray}
Here, $V_{L}$ and $V_{R}$ are the external gate potentials applied to the left- and right-BLGs, respectively, as it is presented in Fig.~\ref{fig:Fig_1}. The time independent term ${\Delta\varphi}_{0}/2$ in Eq.(\ref{Equation_13}) is given as ${\Delta\varphi}_{0}={\varphi}_{L}-\varphi_{R}$ and will play the role of the phase detuning parameter in the expression of the excitonic current. Furthermore, $\varphi_{L}$ and $\varphi_{R}$ are the macroscopic phases of the coherent excitonic condensates in both sides of the tunnel junction. Furthermore, we will define the time-dependent phase difference parameter ${\Delta\varphi}(t)$ as 
\begin{eqnarray}
{\Delta\varphi}(t)=\frac{2e}{\hbar}\int^{t}_{0}dt'V(t'),
\label{Equation_14}
\end{eqnarray}
where $V(t)=(V_{L}(t)-V_{R}(t))/2$. Next, we will consider the tunnelling matrix term $H_{T}$ as the small perturbation turned on adiabatically from $t=-\infty$ and which determines the time evolution of the eigenstates in the Heisenberg picture. Therefore, in the first order in $H_{T}$, we have
\begin{widetext}
	\begin{eqnarray}
I(V,T,t)=\frac{2e}{\hbar}\Im\left[\left(-\frac{i}{\hbar}\right)\int^{t}_{-\infty}d\tau e^{\eta\tau}\langle\left[A^{\dag}(t)e^{i\Delta{\Phi}_{\rm Lower}},H_{T}(\tau)\right]_{-}\rangle_{H'_{0}}+\langle\left[C^{\dag}(t)e^{i\Delta{\Phi}_{\rm Lower}},H_{T}(\tau)\right]_{-}\rangle_{H'_{0}}\right].
\label{Equation_15}
\end{eqnarray}
\end{widetext}
The infinitesimal, positive constant $\eta=0^{+}$ has been introduced in the integral in Eq.(\ref{Equation_15}) in order to assure the convergence of the integral. 
Here, the averages over the commutators are referred to the unperturbed Hamiltonian $H'_{0}=H_{0}+H_{\rm int}$. The time-dependent operators $A^{\dag}(t)$ and $C^{\dag}(t)$ in Eq.(\ref{Equation_15}) have been introduced with respect to the reformulation of the tunnelling Hamiltonian in Eq.(\ref{Equation_7}) in terms of the composite tunnel-operators $A(t), B(t),C(t)$ and $D(t)$. Namely, we write
	\begin{eqnarray}
H_{T}=\left[A^{\dag}(t)+C^{\dag}(t)+B(t)+D(t)\right]e^{i\Phi_{\rm Lower}(t)}+c.c.,
\nonumber\\
\label{Equation_16}
\end{eqnarray}
where 
 	\begin{eqnarray}
 A(t)=\sum_{{\bf{k}},{\bf{p}}}\sum_{\sigma}t^{\ast}_{{\bf{k}},{\bf{p}}}\tilde{c}^{\dag}_{{\bf{p}},\sigma}(t)\tilde{a}_{{\bf{k}},\sigma}(t),
 \nonumber\\
  B(t)=\sum_{{\bf{k}},{\bf{p}}}\sum_{\sigma}t^{\ast}_{{\bf{k}},{\bf{p}}}{d}^{\dag}_{{\bf{p}},\sigma}(t){b}_{{\bf{k}},\sigma}(t),
  \nonumber\\
   C(t)=\sum_{{\bf{k}},{\bf{p}}}\sum_{\sigma}t^{\ast}_{{\bf{k}},{\bf{p}}}\tilde{d}^{\dag}_{{\bf{p}},\sigma}(t)\tilde{b}_{{\bf{k}},\sigma}(t),
   \nonumber\\
    D(t)=\sum_{{\bf{k}},{\bf{p}}}\sum_{\sigma}t^{\ast}_{{\bf{k}},{\bf{p}}}{c}^{\dag}_{{\bf{p}},\sigma}(t){a}_{{\bf{k}},\sigma}(t).
 \label{Equation_17}
 \end{eqnarray}
 It is important to note that the phase factor in Eq.(\ref{Equation_16}) appears after the Rickayzen transformations, given in Eq.(\ref{Equation_12}), above. 
 %
  %
  \subsection{\label{sec:Section_2_3} The normal quasiparticle and excitonic tunneling}
  %
 Next, we calculate the commutators in Eq.(\ref{Equation_15}) and we perform the statistical averaging of four-point correlation functions using the Wick's theorem at the finite temperatures \cite{cite_46}.  
After the whole averaging procedure, we keep only terms responsible for the normal and excitonic currents in the junction. We obtain for the total tunnelling current the following expression
\begin{widetext}
	\begin{eqnarray}
	I(V,T,t)=\frac{2e}{\hbar}\Im\left[\left(-\frac{i}{\hbar}\right)\int^{t}_{-\infty}d\tau e^{\eta\tau}\left(\langle\left[A^{\dag}(t),A(\tau)\right]_{-}\rangle_{H'_{0}}+\langle\left[C^{\dag}(t),C(\tau)\right]_{-}\rangle_{H'_{0}}\right)e^{i\frac{\Delta{\varphi}(t)-\Delta{\varphi}(\tau)}{2}}\right.
	\nonumber\\
	\left.+\langle\left[A^{\dag}(t),B(\tau)\right]_{-}\rangle_{H'_{0}}e^{i\frac{\Delta{\varphi}(t)+\Delta{\varphi}(\tau)}{2}}e^{i\left(\varphi_{L}-\varphi_{R}\right)}
	\right],
	\label{Equation_18}
	\end{eqnarray}
\end{widetext}
where the first term, in the sum in the right-hand side in Eq.(\ref{Equation_18}), gives the normal single-particle tunnelling current and the second term in Eq.(\ref{Equation_18}) is responsible for the excitonic tunnelling. We obtain
	\begin{eqnarray}
&&I(V,T,t)=\Im\left[ e^{i\frac{\Delta{\varphi(t)}}{2}}
\int^{\infty}_{-\infty}dt' e^{\eta(t-t')}\times\right.
\nonumber\\
&&\times\left(e^{-i\frac{\Delta{\varphi(t-t')}}{2}}N(t')
\left.+e^{i\frac{\Delta{\varphi(t-t')}}{2}}F(t')\right)e^{i{\Delta\varphi}_{0}}\right].
\label{Equation_19}
\end{eqnarray}
Here, we have performed a change of variable of integration $t'=t-\tau$ and we have introduced the time dependent functions $N(t)$ and $F(t)$ for the normal and excitonic counterparts, which are defined with the help of the Kadanoff-Baym Green's functions \cite{cite_51}. For the normal function $N(t)$, we have
	\begin{eqnarray}
N(t)=&&-\frac{ie}{\pi^{2}{\hbar^{2}}}\Theta(t)\sum_{{\bf{k}}}\int^{\infty}_{-\infty}\int^{\infty}_{-\infty}d\omega{d\omega'}e^{i(\omega-\omega')}\times
\nonumber\\
&&\times\left[A^{L}_{\tilde{a}}({\bf{k}},\omega)A^{R}_{\tilde{c}}({\bf{k}},\omega')+A^{L}_{\tilde{b}}({\bf{k}},\omega)A^{R}_{\tilde{d}}({\bf{k}},\omega')\right]\times
\nonumber\\
&&\times \left[n^{L}_{F}(\omega)-n^{R}_{F}(\omega')\right]
	\label{Equation_20}
	\end{eqnarray}
and for the excitonic function $F(t)$, we get
	\begin{eqnarray}
F(t)=&&-\frac{ie}{\pi^{2}{\hbar^{2}}}\Theta(t)\sum_{{\bf{k}}}\int^{\infty}_{-\infty}\int^{\infty}_{-\infty}d\omega{d\omega'}e^{i(\omega-\omega')}\times
\nonumber\\
&&\times A^{L}_{\tilde{a}b}({\bf{k}},\omega)A^{R}_{\tilde{c}d}({\bf{k}},\omega')\left[n^{L}_{F}(\omega)-n^{R}_{F}(\omega')\right].
	\label{Equation_21}
\end{eqnarray}
Th function $n^{\ell}_{F}(\omega)=1/(e^{\beta{\omega}}+1)$ ($\ell=L,R$) in Eqs.(\ref{Equation_20}) and (\ref{Equation_21}) is the Fermi-Dirac distribution function. We have performed the Fourier transformation of the Kadanoff's Green's functions \cite{cite_51} in order to obtain the expressions of $N(t)$ and $F(t)$, given in Eqs.(\ref{Equation_20}) and (\ref{Equation_21}) and, also, we have introduced the single-particle spectral functions $A^{\ell}_{X_{\ell}}({\bf{k}},\omega)$ ($X=\tilde{a}, \tilde{b}$ for $\ell=L$ and $X=\tilde{c}, \tilde{d}$ for $\ell=R$) and excitonic spectral functions $A^{L}_{\tilde{a}b}({\bf{k}},\omega)$, $A^{R}_{\tilde{c}d}({\bf{k}},\omega)$. We have supposed in Eqs.(\ref{Equation_20}) and (\ref{Equation_21}) a simple form of the tunnelling probability $t_{{\bf{k}},{\bf{p}}}=\delta_{{\bf{k}},{\bf{p}}}$. For a simple treatment, this approximation is sufficient to consider the excitonic effects. 

The normal and anomalous (or excitonic) spectral functions in the separate BLGs with the presence of the excitonic pairing interaction have been discussed by us in Ref.\onlinecite{cite_38}, where the full four-band theory has been developed without low-energy assumption near the $K$-point in the Brillouine zone. From the form of the total fermionic action, derived there, it follows that the normal spectral functions in different layers in the left-BLG are interconnected:
\begin{eqnarray}
A^{L}_{\tilde{a}}({\bf{k}},\omega)=A^{L}_{b}({\bf{k}},\omega),
\nonumber\\
A^{L}_{\tilde{b}}({\bf{k}},\omega)=A^{L}_{a}({\bf{k}},\omega).
\label{Equation_22}
\end{eqnarray}
The same is true also for the right-BLG system. Moreover, the explicit analytical expressions of the normal spectral functions in bottom layer of the left-BLG are \cite{cite_38}
\begin{eqnarray}
A^{L}_{a}({\bf{k}},\omega)=\sum_{i=1,...4}\alpha^{L}_{i{\bf{k}}}\delta(\omega+\kappa^{L}_{i{\bf{k}}}),
\nonumber\\
A^{L}_{b}({\bf{k}},\omega)=\sum_{i=1,...4}\beta^{L}_{i{\bf{k}}}\delta(\omega+\kappa^{L}_{i{\bf{k}}})
\label{Equation_23}
\end{eqnarray}
and the similar expressions can be written for the right-BLG structure $\ell=R$. The spectral functions in the upper layers in both BLGs could be obtained after the relations in Eq.(\ref{Equation_22}). The ${\bf{k}}$-dependent coefficients $\alpha^{\ell}_{i{\bf{k}}}$ and $\beta^{\ell}_{i{\bf{k}}}$ are given in Ref.\onlinecite{cite_38}. The excitonic energy dispersion parameters $\kappa^{\ell}_{i{\bf{k}}}$ in Eq.(\ref{Equation_23}), which define the band structures of the interacting BLGs in both sides of the tunnel junction, are given by \cite{cite_38}
\begin{widetext}
	
	\begin{eqnarray}
	\kappa^{\ell}_{1,2{\bf{k}}}=-\frac{1}{2}\left[\Delta_{\ell}+\gamma_{1}\pm\sqrt{\left(W_{\ell}-\Delta_{\ell}-\gamma_{1}\right)^{2}+4|\tilde{\gamma}_{{\bf{k}}}|^{2}}\right]+\bar{\mu}_{\ell},
	\nonumber\\
	\kappa^{\ell}_{3,4{\bf{k}}}=-\frac{1}{2}\left[-\Delta_{\ell}-\gamma_{1}\pm\sqrt{\left(W_{\ell}+\Delta_{\ell}+\gamma_{1}\right)^{2}+4|\tilde{\gamma}_{{\bf{k}}}|^{2}}\right]+\bar{\mu}_{\ell},
	\label{Equation_24}
	\end{eqnarray}
	\newline\\
\end{widetext} 
As we have mentioned above, the intralayer and interlayer hopping parameters are supposed the same in both sides of the junction. It is particularly important to underline here the role of the bare chemical potentials $\bar{\mu}_{\ell}$ appearing in the expressions of the band structure parameters in Eq.(\ref{Equation_24}). Indeed, they are playing the role of the exact Fermi energies in the BLG structures as it was pointed out in Ref.\onlinecite{cite_38}. 
For the anomalous (or excitonic) spectral functions in Eq.(\ref{Equation_21}) we have \cite{cite_38}  
\begin{eqnarray}
A^{L}_{\tilde{a}b}({\bf{k}},\omega)=(\gamma_1+\Delta_{L})\sum_{i=1,...4}\gamma^{L}_{i{\bf{k}}}\delta(\omega+\kappa^{L}_{i{\bf{k}}}),
\nonumber\\
A^{R}_{\tilde{c}d}({\bf{k}},\omega)=(\gamma_1+\Delta_{R})\sum_{i=1,...4}\gamma^{R}_{i{\bf{k}}}\delta(\omega+\kappa^{R}_{i{\bf{k}}}),
\label{Equation_25}
\end{eqnarray}
and the parameters $\gamma^{\ell}_{i{\bf{k}}}$ are given in Ref.\onlinecite{cite_38}.

Furthermore, the normal and excitonic tunnelling currents can be simply expressed analytically after the Werthamer spectral decomposition \cite{cite_52} and by supposing simultaneously the case of the constant gate voltage, i.e., $V(t)=V=\text{const}$. We have 
\begin{eqnarray}
 e^{i\frac{{\Delta\varphi}(t)}{2}}=\int^{\infty}_{-\infty}\frac{dE}{2\pi}W^{\ast}(E)e^{iEt},
 \nonumber\\
 	\label{Equation_26}
\end{eqnarray}
and
\begin{eqnarray}
W(E)=W^{\ast}(E)=2\pi\delta\left(E-\frac{e}{\hbar}V\right).
\end{eqnarray}
Here, we have used the definition of the phase difference function in Eq.(\ref{Equation_14}). The total tunnelling current through the tunnel junction will be
\begin{eqnarray}
&&I(V,T,t)=I_{n}(V,T)+I_{\rm Exc}(V,T,t)
 \nonumber\\
=&&\Im\left(N(i\eta+\omega_0)\right)+\Im\left[e^{i(\Delta\varphi_{0}+2\omega_{0}t)}F(i\eta-\omega_{0})\right]
\nonumber\\
=&&I_{\rm qp}(V,T)+I_{\rm J_{1}}(V,T)\sin(\Delta\varphi_{0}+2\omega_{0}t)
\nonumber\\
&&+I_{\rm J_{2}}(V,T)\cos(\Delta\varphi_{0}+2\omega_{0}t),
 	\label{Equation_27}
\end{eqnarray}
where $\omega_0$ is the field-frequency: $\omega_{0}=\frac{e}{\hbar}V$, and the normal tunnelling is equivalent to the single-particle tunnelling term $I_{n}(V,T)=\Im(N(i\eta+\omega_0))$, while the coherent Josephson tunnelling of excitons is given by the last two terms in Eqs.(\ref{Equation_27}). Thus, the total excitonic tunnelling current is 
\begin{eqnarray}
I_{\rm Exc}(V,T,t)=&&I_{\rm J_{1}}(V,T)\sin(\Delta\varphi_{0}+2\omega_{0}t)
\nonumber\\
&&+I_{\rm J_{2}}(V,T)\cos(\Delta\varphi_{0}+2\omega_{0}t),
	\label{Equation_28}
\end{eqnarray}
and we have $I_{\rm J_{1}}(V,T)=\Re(F(i\eta-\omega_{0}))$ and $I_{\rm J_{2}}(V,T)=\Im(F(i\eta-\omega_{0}))$.
 
Finally, after some calculations, we get for the normal tunnelling current the following expression
\begin{widetext}
\begin{eqnarray}
I_{n}(V,T)=-\frac{e}{\pi\hbar^{2}}\left\{\Theta_{\rm T}\left(\Delta_{\omega_{0}},a_{1},b_{1}\right)\sum_{i=1,2}\left[\rho(x_{i})\frac{\alpha^{L}_{1}(x_{i})\alpha^{R}_{2}(x_{i})+\beta^{L}_{1}(x_{i})\beta^{R}_{2}(x_{i})}{|f(x_{i},a_{1},b_{1})|}\left(n^{L}_{F}(-\kappa^{L}_{1}(x_{i}))-n^{R}_{F}\left(-\kappa^{L}_{1}(x_{i})+\omega_{0}\right)\right)\right]\right.
\nonumber\\
\left.+\Theta_{\rm T}\left(\Delta'_{\omega_{0}},c_{1},b_{1}\right)\sum_{i=3,4}\left[\rho(x_{i})\frac{\alpha^{L}_{1}(x_{i})\alpha^{R}_{4}(x_{i})+\beta^{L}_{1}(x_{i})\beta^{R}_{4}(x_{i})}{|f(x_{i},c_{1},b_{1})|}\left(n^{L}_{F}(-\kappa^{L}_{1}(x_{i}))-n^{R}_{F}\left(-\kappa^{L}_{1}(x_{i})+\omega_{0}\right)\right)\right]\right.
\nonumber\\
\left.+\Theta_{\rm T}\left(-\Delta_{\omega_{0}},a_{1},b_{1}\right)\sum_{i=1,2}\left[\rho(x_{i})\frac{\alpha^{L}_{2}(x_{i})\alpha^{R}_{1}(x_{i})+\beta^{L}_{2}(x_{i})\beta^{R}_{1}(x_{i})}{|f(x_{i},a_{1},b_{1})|}\left(n^{L}_{F}(-\kappa^{L}_{2}(x_{i}))-n^{R}_{F}\left(-\kappa^{L}_{2}(x_{i})+\omega_{0}\right)\right)\right]\right.
\nonumber\\
\left.+\Theta_{\rm T}\left(-\Delta'_{\omega_{0}},c_{1},b_{1}\right)\sum_{i=3,4}\left[\rho(x_{i})\frac{\alpha^{L}_{2}(x_{i})\alpha^{R}_{3}(x_{i})+\beta^{L}_{2}(x_{i})\beta^{R}_{3}(x_{i})}{|f(x_{i},c_{1},b_{1})|}\left(n^{L}_{F}(-\kappa^{L}_{2}(x_{i}))-n^{R}_{F}\left(-\kappa^{L}_{2}(x_{i})+\omega_{0}\right)\right)\right]\right.
\nonumber\\
\left.+\Theta_{\rm T}\left(\tilde{\Delta}'_{\omega_{0}},a_{1},d_{1}\right)\sum_{i=5,6}\left[\rho(x_{i})\frac{\alpha^{L}_{3}(x_{i})\alpha^{R}_{2}(x_{i})+\beta^{L}_{3}(x_{i})\beta^{R}_{2}(x_{i})}{|f(x_{i},a_{1},d_{1})|}\left(n^{L}_{F}(-\kappa^{L}_{3}(x_{i}))-n^{R}_{F}\left(-\kappa^{L}_{3}(x_{i})+\omega_{0}\right)\right)\right]\right.
\nonumber\\
\left.+\Theta_{\rm T}\left(\tilde{\Delta}_{\omega_{0}},c_{1},d_{1}\right)\sum_{i=7,8}\left[\rho(x_{i})\frac{\alpha^{L}_{3}(x_{i})\alpha^{R}_{4}(x_{i})+\beta^{L}_{3}(x_{i})\beta^{R}_{4}(x_{i})}{|f(x_{i},c_{1},d_{1})|}\left(n^{L}_{F}(-\kappa^{L}_{3}(x_{i}))-n^{R}_{F}\left(-\kappa^{L}_{3}(x_{i})+\omega_{0}\right)\right)\right]\right.
\nonumber\\
\left.+\Theta_{\rm T}\left(-\tilde{\Delta}'_{\omega_{0}},a_{1},d_{1}\right)\sum_{i=5,6}\left[\rho(x_{i})\frac{\alpha^{L}_{4}(x_{i})\alpha^{R}_{1}(x_{i})+\beta^{L}_{4}(x_{i})\beta^{R}_{1}(x_{i})}{|f(x_{i},a_{1},d_{1})|}\left(n^{L}_{F}(-\kappa^{L}_{4}(x_{i}))-n^{R}_{F}\left(-\kappa^{L}_{4}(x_{i})+\omega_{0}\right)\right)\right]\right.
\nonumber\\
\left.+\Theta_{\rm T}\left(-\tilde{\Delta}_{\omega_{0}},c_{1},d_{1}\right)\sum_{i=7,8}\left[\rho(x_{i})\frac{\alpha^{L}_{4}(x_{i})\alpha^{R}_{3}(x_{i})+\beta^{L}_{4}(x_{i})\beta^{R}_{3}(x_{i})}{|f(x_{i},c_{1},d_{1})|}\left(n^{L}_{F}(-\kappa^{L}_{4}(x_{i}))-n^{R}_{F}\left(-\kappa^{L}_{4}(x_{i})+\omega_{0}\right)\right)\right]\right\}.
\nonumber\\
\label{Equation_29}
\end{eqnarray}
	\newline\\
\end{widetext}
Here, we have introduced a product-$\Theta$ function $\Theta_{T}(x,y,z)$, in the following way
\begin{widetext}
\begin{eqnarray}
\Theta_{T}(x,y,z)=\theta(x)\theta(x^{2}-y^{2}-z^{2})\theta\left[(x^{2}-y^{2}-z^{2})^{2}-4y^{2}z^{2}\right]\theta(\sqrt{x^{4}+4y^{2}z^{2}}-y^{2}-z^{2}),
\label{Equation_30}
\end{eqnarray}
\newline\\
\end{widetext}
where each multiplier, in the total product, is given as a single $\theta$-Heaviside step function. Furthermore, the interaction dependent parameters $a_{1}, b_{1},c_{1}$ and $d_{1}$, in Eq.(\ref{Equation_29}), are defined as 
  \begin{eqnarray}
  a_{1}=W_{R}-\Delta_{R}-\gamma_{1},
  \nonumber\\
  b_{1}=W_{L}-\Delta_{L}-\gamma_1,
  \nonumber\\
  c_{1}=W_{R}+\Delta_{R}+\gamma_{1},
   \nonumber\\
  d_{1}=W_{L}+\Delta_{L}+\gamma_1.
  \label{Equation_31}
  \end{eqnarray}
  Next, the frequency-dependent parameters ${\Delta}_{\omega_{0}}, {\Delta}'_{\omega_{0}}, \tilde{\Delta}_{\omega_{0}}$ and $\tilde{\Delta}'_{\omega_{0}}$, introduced in Eq.(\ref{Equation_29}), form a detuning matrix 
  \begin{eqnarray}
  \hat{\Delta}_{\omega}=\left(
  \begin{array}{ccrr}
 {\Delta}_{\omega_{0}} & {\Delta}'_{\omega_{0}} \\
  \tilde{\Delta}'_{\omega_{0}}  &\tilde{\Delta}_{\omega_{0}}
 \end{array}
  \right),
  \label{Equation_32}
  \end{eqnarray}
  and for each component we have
  \begin{eqnarray}
  {\Delta}_{\omega_{0}}=2(\omega_{0}-\omega'_0),
  \nonumber\\
  \tilde{{\Delta}}_{\omega_{0}}=2(\omega_{0}-\tilde{\omega}'_0),
  \nonumber\\
  {\Delta}'_{\omega_{0}}=2(\omega_{0}-\omega''_0),
  \nonumber\\
  \tilde{{\Delta}}'_{\omega_{0}}=2(\omega_{0}-\tilde{\omega}''_0),
    \label{Equation_33}
  \end{eqnarray}
  where the detuning frequencies $\omega'_0,  \tilde{\omega}'_0, \omega''_0$ and $\tilde{\omega}''_0$ depend explicitly on the difference between the Fermi energies in different sides of the tunnel junction. Namely, we get 
   \begin{eqnarray}
  \omega'_0=\bar{{\mu}}_{R}-\bar{\mu}_{L}-\frac{1}{2}(\Delta_{R}-\Delta_{L}),
  \nonumber\\
  \tilde{\omega}'_0=\bar{{\mu}}_{R}-\bar{\mu}_{L}+\frac{1}{2}(\Delta_{R}-\Delta_{L}),
  \nonumber\\
  \omega''_0=\bar{{\mu}}_{R}-\bar{\mu}_{L}+\frac{1}{2}(\Delta_{R}+\Delta_{L}+2\gamma_1),
  \nonumber\\
  \tilde{\omega}''_0=\bar{{\mu}}_{R}-\bar{\mu}_{L}-\frac{1}{2}(\Delta_{R}+\Delta_{L}+2\gamma_1).
   \label{Equation_34}
  \end{eqnarray}
Next, eight parameters $x_{i}$ with $i=1,...8$, have been introduced in Eq.(\ref{Equation_29}), which are defined as 
\begin{eqnarray}
x_{1,2}=\pm X({\Delta}_{\omega_{0}},a_{1},b_{1}),
  \nonumber\\ 
x_{3,4}=\pm X( {\Delta}'_{\omega_{0}}, b_{1},c_{1}),  
  \nonumber\\
x_{5,6}=\pm X(\tilde{{\Delta}}'_{\omega_{0}}, a_{1},d_{1}),
  \nonumber\\
x_{7,8}=\pm X(\tilde{{\Delta}}_{\omega_{0}}, c_{1},d_{1})
\label{Equation_35}
\end{eqnarray}
whereas, the function $X(x,y,z)$ is defined as $X(x,y,z)=(4\gamma_{0}|x|)^{-1}\sqrt{\left(x^{2}-y^{2}-z^{2}\right)^{2}-4y^{2}z^{2}}$. The density of states (DOS) function $\rho(x)$ in Eq.(\ref{Equation_29}) appears after transforming the ${\bf{k}}$-summation in Eq.(\ref{Equation_20}) into the integration over the continuous variable, i.e., $\sum_{\mathbf{k}}\ldots=\int dx\rho(x)...$. The DOS, in the non-interacting graphene layer, is defined as
\begin{eqnarray}
\rho(x)=\sum_{{\bf{k}}}\delta(x-\gamma_{{\bf{k}}}),
\label{Equation_35}
\end{eqnarray}
where $\gamma_{{\bf{k}}}$ is the band dispersion in the non-interacting single graphene sheet, i.e.,
\begin{eqnarray}
\gamma_{\bf{k}}=e^{-ik_{x}d}+2\exp{i\frac{k_{x}d}{2}}\cos{\frac{\sqrt{3}}{2}k_{y}d}.
\label{Equation_36}
\end{eqnarray}
The parameter $d$, in Eq.(\ref{Equation_36}), refers to the carbon-carbon distance in the graphene layers. 
Beyond the Dirac's approximation, the DOS can be analytically expressed \cite{cite_39, cite_53}  as
\begin{eqnarray}
\footnotesize
\arraycolsep=0pt
\medmuskip = 0mu\rho(x)=\frac{2|x|}{\pi^{2}|\gamma_{0}|^{2}}
\left\{
\begin{array}{cc}
\displaystyle  & \frac{1}{\sqrt{\Lambda\left(|{x}/{\gamma_0}|\right)}}{\mathbf{K}}\left[\frac{4|x/\gamma_0|}{\Lambda\left(|{x}/{\gamma_0}|\right)}\right],  \ \ \  0<|x|<\gamma_0,
\newline\\
\displaystyle  & \frac{1}{\sqrt{4|{{x}/{\gamma_0}}|}}{\mathbf{K}}\left[\frac{\Lambda\left(|x/\gamma_0|\right)}{4|{x}/{\gamma_0}|}\right],  \ \ \ \gamma_0<|x|<3\gamma_0,
\end{array}\right.
\nonumber\\
\label{Equation_37}
\end{eqnarray}
where ${\mathbf{K}}(x)$ is the Elliptic integral of the first kind \cite{cite_54}  ${\mathbf{K}}(x)=\int^{\pi/2}_{0}dt/\sqrt{1-x^{2}\sin^{2}t}$ .
The function $\Lambda(x)$, in Eq.(\ref{Equation_37}), is given by \cite{cite_53} 
\begin{eqnarray}
\Lambda(x)=\left(1+x\right)^{2}-\frac{\left(x^{2}-1\right)^{2}}{4}.
\label{Equation_38}
\end{eqnarray}
Furthermore, the functions $f(x,y,z)$ in the denominators in Eq.(\ref{Equation_29}) read as:
\begin{eqnarray}
f(x,y,z)=\frac{2x\gamma^{2}_{0}}{\sqrt{y^{2}+4x^{2}\gamma^{2}_{0}}}+\frac{2x\gamma^{2}_{0}}{\sqrt{z^{2}+4x^{2}\gamma^{2}_{0}}}.
\label{Equation_39}
\end{eqnarray}

Concerning the excitonic part of the tunnelling current, the Josephson current $I_{J_1}(V,T)$, in Eq.(\ref{Equation_28}), is given as:
\begin{eqnarray}
I_{J_1}(V,T)=\frac{e}{\pi^{2}\hbar^{2}}{\rm P.V.}\sum_{{\bf{k}}}\int^{\infty}_{-\infty}\int^{\infty}_{-\infty}d\omega d\omega'\times
\nonumber\\
\times\frac{A^{L}_{\tilde{a}b}({\bf{k}},\omega)A^{R}_{\tilde{c}d}({\bf{k}},\omega')}{\omega-\omega'-\omega_0}\left(n^{L}_{F}(\omega)-n^{R}_{F}(\omega')\right).
	\label{Equation_40}
	\end{eqnarray}
Furthermore, we will calculate numerically the principal value ${\rm P.V.}$, in Eq.(\ref{Equation_40}), by using the explicit expressions of the excitonic spectral functions, given in Eq.(\ref{Equation_25}). 
For the Josephson current $I_{J_2}(V,T)$, we get
\begin{widetext}
\begin{eqnarray}
I_{J_2}(V,T)=-\frac{e}{\pi\hbar^{2}}(\Delta_{L}+\gamma_1)(\Delta_{R}+\gamma_1)\sum_{{\bf{k}}}\sum_{i,j=1,...4}\gamma^{L}_{i{\bf{k}}}\gamma^{R}_{j{\bf{k}}}\delta\left(-\kappa^{L}_{i{\bf{k}}}+\kappa^{R}_{j{\bf{k}}}-\omega_{0}\right)\left(n^{L}_{F}(-\kappa^{L}_{i{\bf{k}}})-n^{R}_{F}(-\kappa^{L}_{i{\bf{k}}}-\omega_{0})\right).
\label{Equation_41}
\end{eqnarray}
\end{widetext}
Again, transforming the summation over the wave vectors into the integration, we obtain for the Josephson current $I_{J_{2}}(V,T)$:
\begin{widetext}
	\begin{eqnarray}
	I_{J_{2}}(V,T)&&=-\frac{e}{\pi\hbar^{2}}(\Delta_{L}+\gamma_1)(\Delta_{R}+\gamma_1)\times
	\nonumber\\
	&&\times\left\{\Theta_{\rm T}\left(\Delta_{\omega_{0}},a_{1},b_{1}\right)\sum_{i=1,2}\left[\rho(x_{i})\frac{\gamma^{L}_{1}(x_{i})\gamma^{R}_{2}(x_{i})}{|f(x_{i},a_{1},b_{1})|}\left(n^{L}_{F}(-\kappa^{L}_{1}(x_{i}))-n^{R}_{F}\left(-\kappa^{L}_{1}(x_{i})-\omega_{0}\right)\right)\right]\right.
	\nonumber\\
	&&\left.+\Theta_{\rm T}\left(\Delta'_{\omega_{0}},c_{1},b_{1}\right)\sum_{i=3,4}\left[\rho(x_{i})\frac{\gamma^{L}_{1}(x_{i})\gamma^{R}_{4}(x_{i})}{|f(x_{i},c_{1},b_{1})|}\left(n^{L}_{F}(-\kappa^{L}_{1}(x_{i}))-n^{R}_{F}\left(-\kappa^{L}_{1}(x_{i})-\omega_{0}\right)\right)\right]\right.
	\nonumber\\
	&&\left.+\Theta_{\rm T}\left(-\Delta_{\omega_{0}},a_{1},b_{1}\right)\sum_{i=1,2}\left[\rho(x_{i})\frac{\gamma^{L}_{2}(x_{i})\gamma^{R}_{1}(x_{i})}{|f(x_{i},a_{1},b_{1})|}\left(n^{L}_{F}(-\kappa^{L}_{2}(x_{i}))-n^{R}_{F}\left(-\kappa^{L}_{2}(x_{i})-\omega_{0}\right)\right)\right]\right.
	\nonumber\\
	&&\left.+\Theta_{\rm T}\left(-\Delta'_{\omega_{0}},c_{1},b_{1}\right)\sum_{i=3,4}\left[\rho(x_{i})\frac{\gamma^{L}_{2}(x_{i})\gamma^{R}_{3}(x_{i})}{|f(x_{i},c_{1},b_{1})|}\left(n^{L}_{F}(-\kappa^{L}_{2}(x_{i}))-n^{R}_{F}\left(-\kappa^{L}_{2}(x_{i})-\omega_{0}\right)\right)\right]\right.
	\nonumber\\
	&&\left.+\Theta_{\rm T}\left(\tilde{\Delta}'_{\omega_{0}},a_{1},d_{1}\right)\sum_{i=5,6}\left[\rho(x_{i})\frac{\gamma^{L}_{3}(x_{i})\gamma^{R}_{2}(x_{i})}{|f(x_{i},a_{1},d_{1})|}\left(n^{L}_{F}(-\kappa^{L}_{3}(x_{i}))-n^{R}_{F}\left(-\kappa^{L}_{3}(x_{i})-\omega_{0}\right)\right)\right]\right.
	\nonumber\\
	&&\left.+\Theta_{\rm T}\left(\tilde{\Delta}_{\omega_{0}},c_{1},d_{1}\right)\sum_{i=7,8}\left[\rho(x_{i})\frac{\gamma^{L}_{3}(x_{i})\gamma^{R}_{4}(x_{i})}{|f(x_{i},c_{1},d_{1})|}\left(n^{L}_{F}(-\kappa^{L}_{3}(x_{i}))-n^{R}_{F}\left(-\kappa^{L}_{3}(x_{i})-\omega_{0}\right)\right)\right]\right.
	\nonumber\\
	&&\left.+\Theta_{\rm T}\left(-\tilde{\Delta}'_{\omega_{0}},a_{1},d_{1}\right)\sum_{i=5,6}\left[\rho(x_{i})\frac{\gamma^{L}_{4}(x_{i})\gamma^{R}_{1}(x_{i})}{|f(x_{i},a_{1},d_{1})|}\left(n^{L}_{F}(-\kappa^{L}_{4}(x_{i}))-n^{R}_{F}\left(-\kappa^{L}_{4}(x_{i})-\omega_{0}\right)\right)\right]\right.
	\nonumber\\
	&&\left.+\Theta_{\rm T}\left(-\tilde{\Delta}_{\omega_{0}},c_{1},d_{1}\right)\sum_{i=7,8}\left[\rho(x_{i})\frac{\gamma^{L}_{4}(x_{i})\gamma^{R}_{3}(x_{i})}{|f(x_{i},c_{1},d_{1})|}\left(n^{L}_{F}(-\kappa^{L}_{4}(x_{i}))-n^{R}_{F}\left(-\kappa^{L}_{4}(x_{i})-\omega_{0}\right)\right)\right]\right\}.
	\label{Equation_42}
	\end{eqnarray}
	\newline\\
\end{widetext}
In the following section we will analyse numerically the obtained expressions for the normal and excitonic tunneling currents in the system.
 \section{\label{sec:Section_3} The numerical results and discussions}
 %
 \subsection{\label{sec:Section_3_1} The normal quasiparticle tunneling}
\begin{figure}
	\begin{center}
		\includegraphics[width=220px,height=240px]{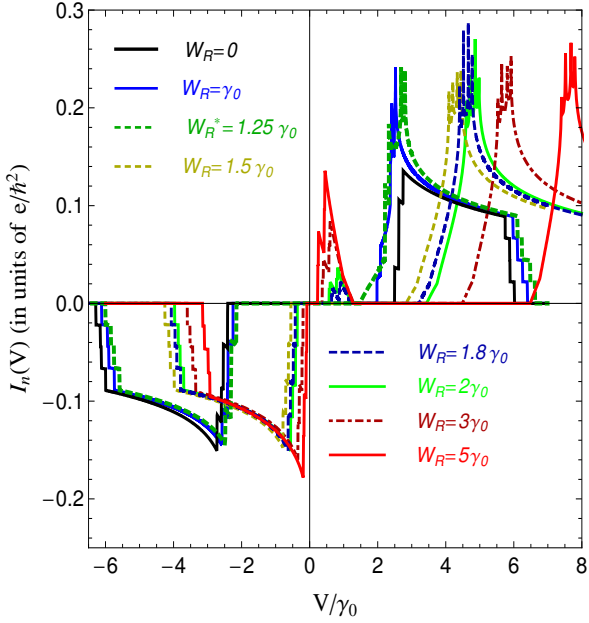}
		\caption{\label{fig:Fig_2}(Color online) The $I-V$ characteristic of the normal quasiparticle tunnelling current, given in Eq.(\ref{Equation_29}), for $W_L=0.5\gamma_0$ and for different values of $W_R$. The low energy quasiparticle tunneling formation, coming from the condensate states is shown at the large values of the interaction parameter $W_R$.}.
	\end{center}
\end{figure}

\begin{figure}
	\begin{center}
		\includegraphics[width=220px,height=240px]{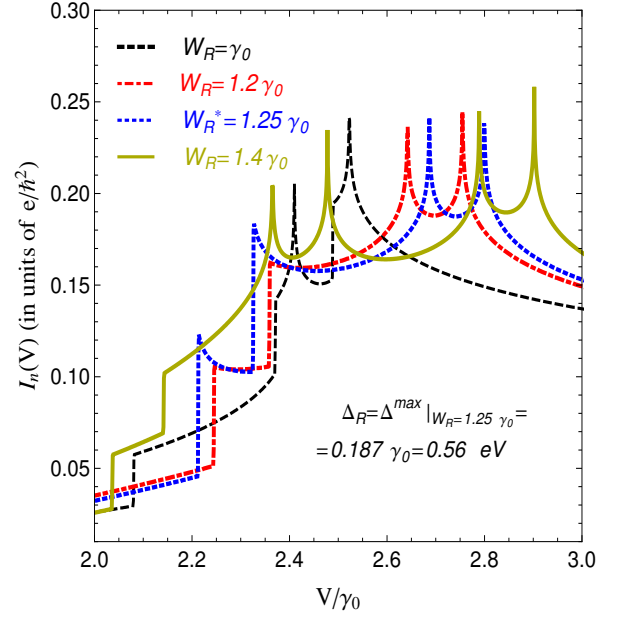}
		\caption{\label{fig:Fig_3}(Color online) The four-bound peaks formations in the normal tunnelling current in the BLG/I/BLG system, given in Eq.(\ref{Equation_29}). The interaction parameter in the left-BLG is fixed at the value $W_L=0.5\gamma_0$ and different values of $W_R$ are considered. The maximal value of the excitonic gap parameter is shown in the figure, corresponding to the value $W^{\ast}_R=1.25\gamma_0$.}.
	\end{center}
\end{figure}

\begin{figure}
	\begin{center}
		\includegraphics[width=230px,height=620px]{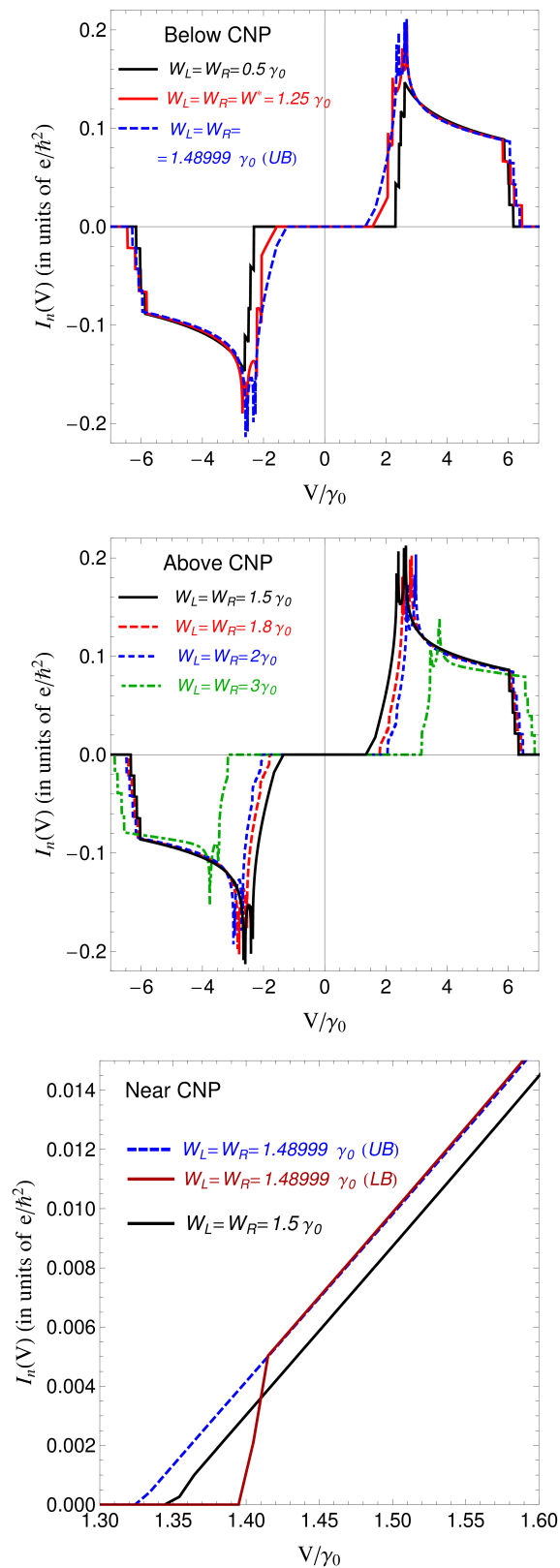}
		\caption{\label{fig:Fig_4}(Color online) The $I-V$ characteristic of the normal quasiparticle tunnelling current, given in Eq.(\ref{Equation_29}) for equal values of the interlayer Coulomb interaction parameters in both sides of the junction: $W_L=W_{R}$. The values of interaction parameters below the CNP point (upper panel), above the CNP point (middle panel) and in the vicinity of $W^{C}$ are considered in the picture.}.
	\end{center}
\end{figure}

 %
In Fig.~\ref{fig:Fig_2}, we have studied the evolution of the normal quasiparticle tunnelling current (given in Eq.(\ref{Equation_29})) as a function of the applied gate voltage $V$ and for two different limits of the right-BLG interlayer Coulomb interaction parameter $W_R$: the weak and intermediate regime, starting from $W_R=0$ (solid black curve), $W_R=\gamma_0$ (solid blue curve), $W^{\ast}_R=1.25\gamma_0$ (bold-dashed darker-green curve) and $W_R=1.5\gamma_0$ (dashed darker-yellow curve) (the value $W_R=1.5\gamma_0$ is chosen very close to the CNP value $W^{C}_{R}=1.48999\gamma_0$) and the other, high interaction limit, when $W_R=1.8\gamma_0$ (dashed darker blue curve), $W_R=2\gamma_0$ (solid green curve), $W_R=3\gamma_0$ (dot-dashed darker red curve) and $W_R=5\gamma_0$ (solid red curve). We denoted by $W^{\ast}_{R}$ the value of the interlayer Coulomb interaction parameter at which the excitonic gap parameter is maximal: $\Delta_{R}=\Delta^{\rm max}_{R}$. The interlayer interaction parameter in the left-BLG is fixed at the value $W_L=0.5\gamma_0$, for all curves in Fig.~\ref{fig:Fig_2}. We see first of all, in Fig.~\ref{fig:Fig_2}, that the normal quasiparticle tunneling through the BLG/I/BLG heterojunction, accompanied with the excitonic pair formations in the system, is a threshold process, and the threshold frequency $\omega_0=(e/\hbar)V$ of the external field depends on the relative values of the Coulomb interaction parameters $W_L$ and $W_R$ at different sides of the construction. We also see that when augmenting the parameter $W_R$ in the interval $W_R\in[0,1.5\gamma_0]$, the curves, corresponding to the positive part of the current function $I_{n}(V,T)$, are shifting into left and the intensity of curves is increased. In turn, the threshold values of the normal tunneling current are also shifting left. Starting from the upper bound (UB) critical CNP value of $W_R$ (i.e., $W_R\geq W^{C}_{R}(UB)$), related to the upper bound solution of the chemical potential in the BLG (see in Ref.\onlinecite{cite_38}), the normal tunneling current is shifting right, on the $V$-axis. Nevertheless, the same is not true for the negative part of the tunnel current. We see, in Fig.~\ref{fig:Fig_2}, that all curves of the negative part of the quasiparticle tunneling current are displacing to right when increasing the interlayer Coulomb interaction parameter in the interval $W_R\in[0, 5\gamma_0]$. Only a large jump of the threshold voltages occurs when passing across the lower bound (LB) CNP value $W^{C}_{R}(LB)$, related to the lower bound chemical potential at the CNP. We observe also that for a very large disbalance between the values of the parameters $W_L$ and $W_R$, the additional low-frequency peaks appear in the positive part of the current spectrum, and the threshold frequency values of $V$ are gradually decreasing in these cases. We see also that the amplitudes of the low-frequency peaks are increasing with $W_R$. It is interesting to note that for $W_R=5\gamma_0$ (see the solid red curve in Fig.~\ref{fig:Fig_2}), the threshold frequency $\omega_0$ in the positive part of the normal current is of order of $\omega_0\sim 2\gamma_1=0.256\gamma_0=0.76$ eV. 
%
%
Another important observation in Fig.~\ref{fig:Fig_2} is related to the formation of the 4-peak like structures in the spectrum of the normal tunneling current (this is due to the strong excitonic excitations in the 4-band structure of the BLGs). Namely, for small values of $W_R$, the spectrum of the normal current is step-wise, which furthermore transforms in to the 4-peak structure at the intermediate values of $W_R$ ($W_R=1.2\gamma_0$, $W^{\ast}_R=1.25\gamma_0$, $W_R=1.5\gamma_0$ and $W_R=1.8\gamma_0$). This is more apparent in Fig.~\ref{fig:Fig_3}, where we have chosen very close values of $W_R$, in order to demonstrate the gradual formation of the current 4-peak structure. 
%
%
We see in Fig.~\ref{fig:Fig_3} that for $W_R=\gamma_0$ the structure of the tunneling current is half-stepwise with the well formed 2 peaks in the excitation spectrum and at the relatively high values of the gate potential. When slightly augmenting $W_R$ (see the curves at $W_R=1.2\gamma_0$ and $W^{\ast}_R=1.25\gamma_0$), the current steps become more pronounced and, at the value $W^{\ast}_{R}=1.25\gamma_0$, the resonant tunneling peaks appear at the place of the current steps. Remember that at $W^{\ast}_{R}=1.25\gamma_0$ the excitonic gap parameter attains its maximum value (see in Ref.\onlinecite{cite_38}). Moreover, for higher values of $W_R$, the 4-peak structure remains present and, additionally, the low-frequency resonant tunneling peaks appear in the electron tunneling spectrum. We relate the 4-peak structure to the high energy strong resonant tunneling of single electrons, and the presence of the very large tunneling threshold is a direct consequence of the EI state in the bilayer graphenes. Thus, in order to do the tunneling, the electrons must break their contribution to the excitonic insulator state. Contrary, the additional low-frequency peaks at the large values of the parameter $W_R$, are related to the excitonic condensate states in the BLGs. This is the manifestation of tunneling coming from the coherent excitonic condensates states in the system. Vis a vis the high interaction values of $W_R$, accompanying the low-frequency peaks, coming from the excitonic condenates states, we can conclude that the excitonic insulator state and the excitonic condensates states in the system are not identical. These are two different states of matter and the EI state does not survives at the high values of $W_R$, considered in Fig.~\ref{fig:Fig_2}. This observation is in complete agreement with the ideas retrieved in Ref.\onlinecite{cite_38}.  This statement is also in agreement with the recent work in Ref.\onlinecite{cite_44}, where it has been shown that the excitonic Josephson current becomes extremely small before the EI state breaks down. A detailed analysis of the role of the CNP point $W^{C}=1.48999\gamma_0$, for the case of interaction balanced BLGs, is given in Fig.~\ref{fig:Fig_4}. Particularly, in the upper panel in Fig.~\ref{fig:Fig_4}, we consider the equal values of $W_L$ and $W_R$, below the upper bound critical value $W^{C}_{L}(UB)=W^{C}_{R}(UB)\equiv W^{C}(UB)$ (the latest was also considered in the figure). We observe that the tunneling spectrum is perfectly symmetric in this case with respect to the origin. When augmenting the interaction parameter up to the upper bound solution at the CNP point, the positive part of the tunneling spectrum is shifted left (see the upper panel in Fig.~\ref{fig:Fig_4}), while for higher values of $W_R$ ($W_R=W_L \geq W^{C}(UB)$), the spectrum is shifted to right (this behavior is presented in the middle panel, in Fig.~\ref{fig:Fig_4}). 
%
%
\begin{figure}
	\begin{center}
		\includegraphics[width=230px,height=240px]{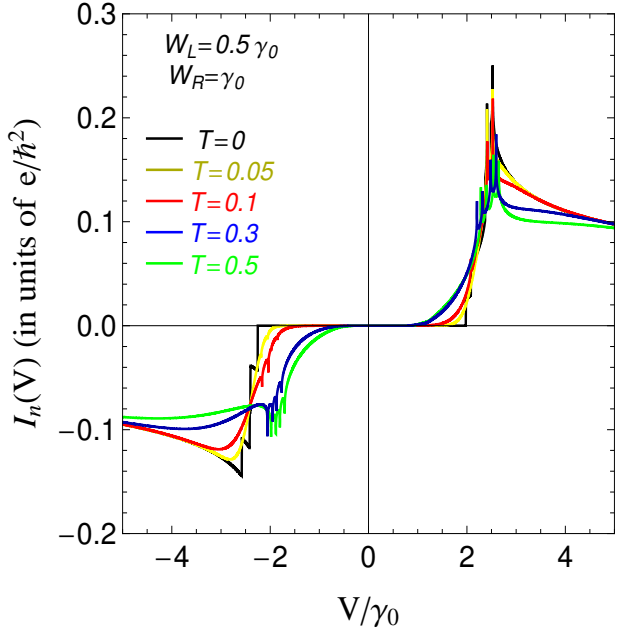}
		\caption{\label{fig:Fig_5}(Color online) Temperature dependence of the normal quasiparticle tunnelling current in the case of the non-equal interlayer Coulomb interactions: $W_L=0.5\gamma_0$ and $W_R=\gamma_0$.}.
	\end{center}
\end{figure} 
%
In the lower panel, in Fig.~\ref{fig:Fig_4}, we have presented the smooth passage of the tunneling spectrum when crossing the CNP point. Both, lower bound and upper bound curves of the tunneling current have been considered there. It is worth to mention that the electronic band structure of the BLGs is doubly degenerated at $W^{C}$ because of the chemical potential solutions in the system (see in Ref.\onlinecite{cite_38}). We see in the bottom panel in Fig.~\ref{fig:Fig_4} that the left outermost curve is the UB normal tunneling current in the system. When passing from LB to UB, the tunneling current spectrum still shifted to the left, while for $W>W^{C}(UB)$ the spectrum is transferring to the right. In Fig.~\ref{fig:Fig_5}, we have presented the temperature dependence of the normal tunneling current for a special antisymmetric choice of the Coulomb interaction parameters $W_L$ and $W_R$: $W_L=0.5\gamma_0$ and $W_R=\gamma_0$. As it is clear from the picture, the amplitude of the normal current decreases with increasing the temperature and also leads to the partial suppression of the threshold values of gate voltage, thus promoting a more flexible tunneling of the normal electrons.    
%
\subsection{\label{sec:Section_3_2} The excitonic Josephson tunneling}
\begin{figure}
	\begin{center}
		\includegraphics[width=200px,height=600px]{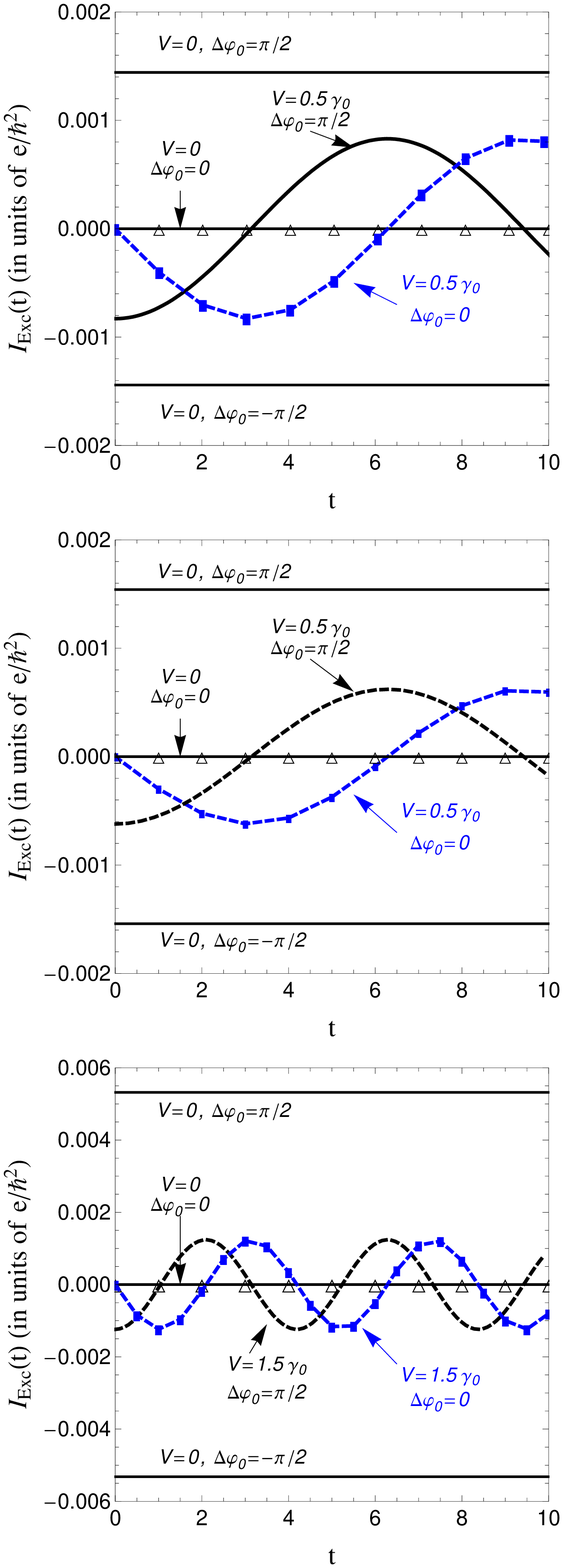}
		\caption{\label{fig:Fig_6}(Color online) The time dependence of the total excitonic Josephson tunnelling current, given in Eq.(\ref{Equation_28}), for different values of the right-BLG interlayer interaction parameter $W_R$ ($W_R=0, 0.5\gamma_0$ and $\gamma_0$, from top to bottom) and for a fixed value of the left-BLG interlayer interaction $W_L=0.5\gamma_0$. Different combinations of the applied gate potential $V$ and the condensate phase difference ${\Delta\varphi}_{0}$ are considered in the panels. The temperature is set to zero, in all panels.}.
	\end{center}
\end{figure}

\begin{figure}
	\begin{center}
		\includegraphics[width=230px,height=210px]{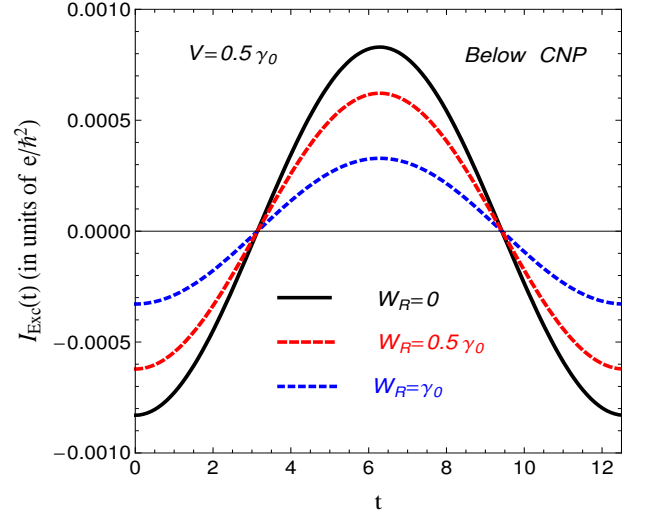}
		\caption{\label{fig:Fig_8}(Color online) The time dependence of the total excitonic Josephson tunnelling current, given in Eq.(\ref{Equation_28}), for $V=0.5\gamma_0$ and for different values of the right-BLG interlayer Coulomb interaction parameter $W_R<W^{C}$. The parameter $W_L$ is fixed at the value $W_L=0.5\gamma_0$ and the condensates phase difference is fixed at the value ${\Delta\varphi}_{0}=\pi/2$. Zero temperature case is considered here.}.
	\end{center}
\end{figure}

\begin{figure}
	\begin{center}
		\includegraphics[width=230px,height=210px]{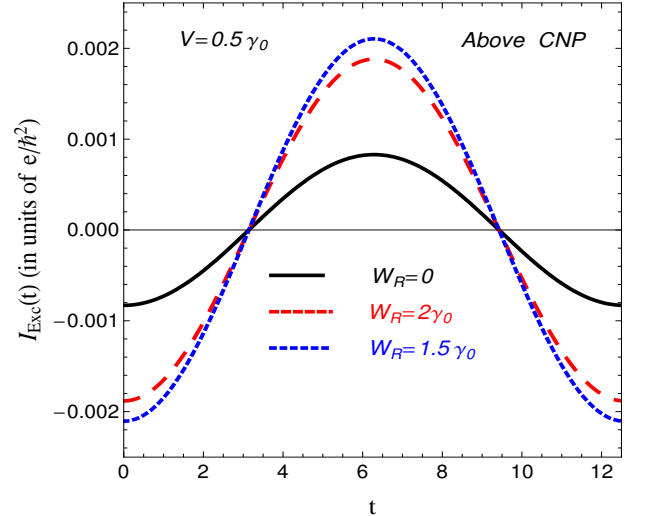}
		\caption{\label{fig:Fig_9}(Color online) The time dependence of the total excitonic Josephson tunnelling current, given in Eq.(\ref{Equation_28}), for $V=0.5\gamma_0$ and for different values of the right-BLG interlayer Coulomb interaction parameter $W_R>W^{C}$. The value $W_R=0$ is also considered as a reference. The parameter $W_L$ is fixed at the value $W_L=0.5\gamma_0$ and the condensates phase difference is fixed at ${\Delta\varphi}_{0}=\pi/2$. Zero temperature case is considered here.}.
	\end{center}
\end{figure}

\begin{figure}
	\begin{center}
		\includegraphics[width=240px,height=450px]{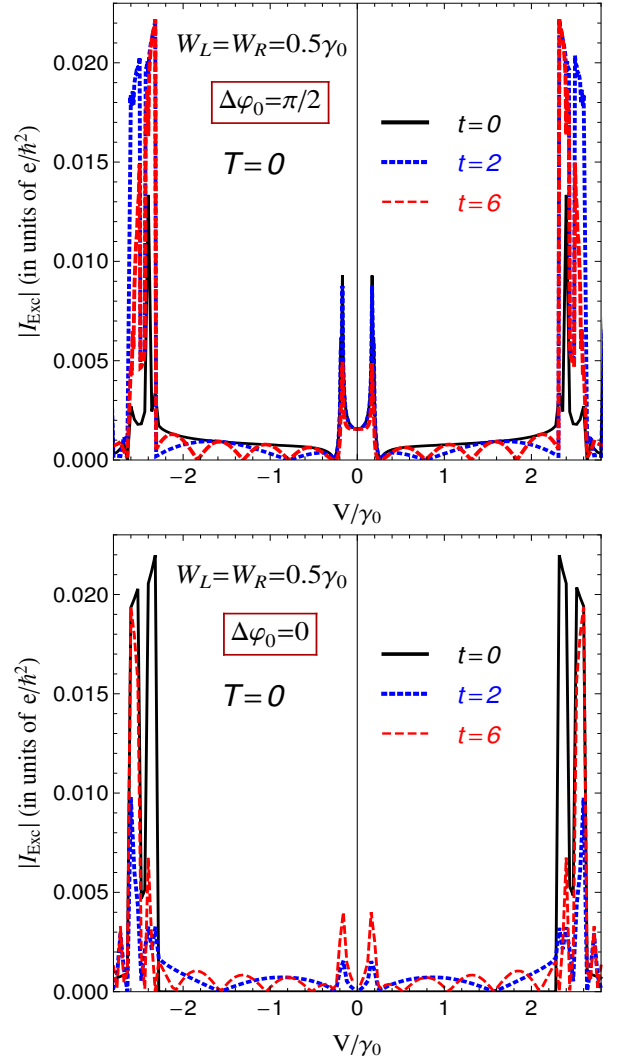}
		\caption{\label{fig:Fig_10}(Color online) The time evolution of the total excitonic Josephson tunnelling current, given in Eq.(\ref{Equation_28}), for the case of $W_L=W_R=0.5\gamma_0$. The condensates phase difference is fixed at the value $\Delta\varphi_{0}=\pi/2$, in the upper panel, and $\Delta\varphi_{0}=0$, in the lower panel. The temperature is set at zero.}.
	\end{center}
\end{figure}

\begin{figure}
	\begin{center}
		\includegraphics[width=240px,height=240px]{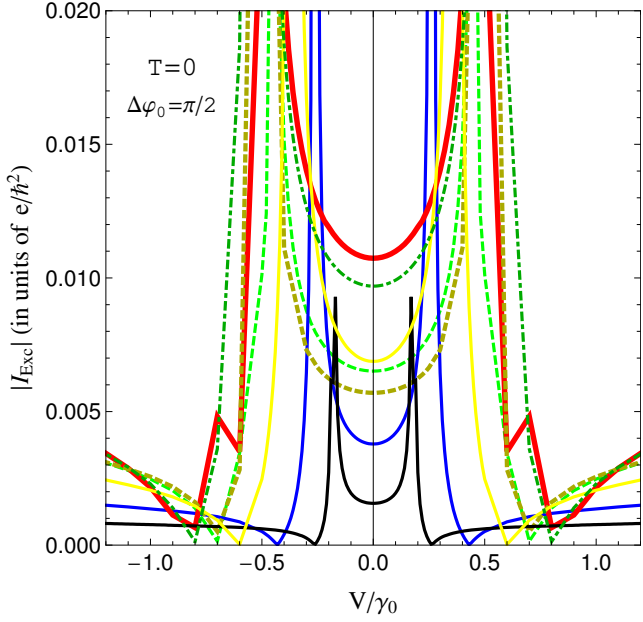}
		\caption{\label{fig:Fig_11}(Color online) The $I-V$ characteristic of the excitonic Josephson tunneling current, given in Eq.(\ref{Equation_28}), for the equal values of the interlayer Coulomb interaction parameters $W_L=W_R$, below the lower bound critical CNP value $W^{C}_{LB}$. The lower bound critical current is shown in dashed dark-yellow curve. The temperature is set at zero.}.
	\end{center}
\end{figure}

\begin{figure}
	\begin{center}
		\includegraphics[width=240px,height=240px]{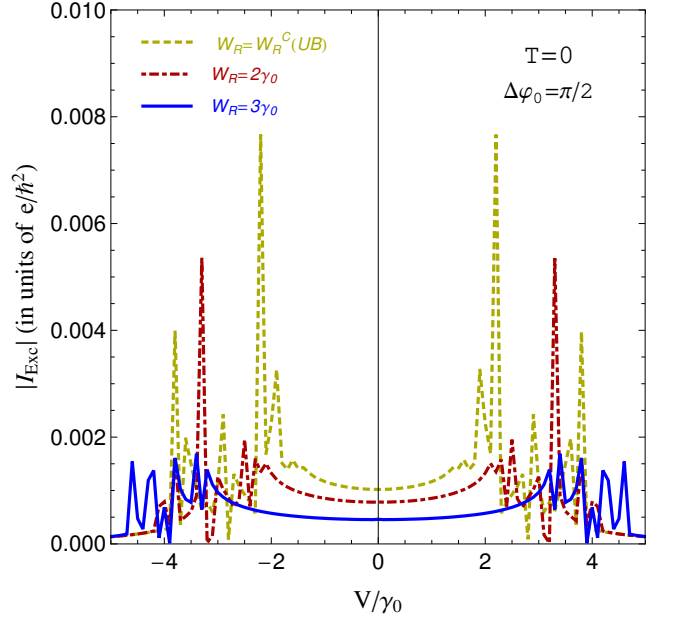}
		\caption{\label{fig:Fig_12}(Color online) The $I-V$ characteristic of the excitonic Josephson tunneling current, given in Eq.(\ref{Equation_28}), for the equal values of the interlayer Coulomb interaction parameters $W_L=W_R$, above the critical upper bound CNP value $W^{C}_{UB}$, presented in dashed dark-yellow curve. The temperature is set at zero.}.
	\end{center}
\end{figure}

\begin{figure}
	\begin{center}
		\includegraphics[width=240px,height=240px]{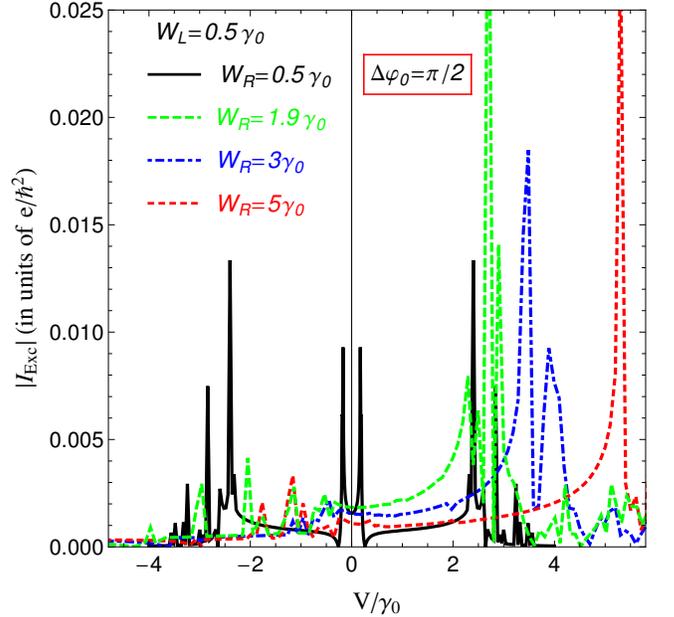}
		\caption{\label{fig:Fig_13}(Color online) The $I-V$ characteristic of the excitonic tunneling current, given in Eq.(\ref{Equation_28}), for $W_L=0.5\gamma_0$ and for different values of $W_R$. The temperature is set at zero.}.
	\end{center}
\end{figure}

\begin{figure}
	\begin{center}
		\includegraphics[width=240px,height=240px]{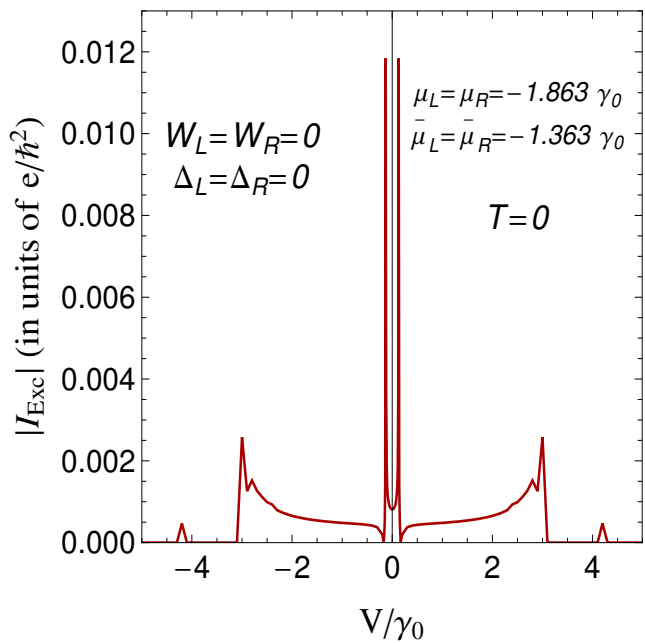}
		\caption{\label{fig:Fig_14}(Color online) The $I-V$ characteristic of the excitonic tunneling current, given in Eq.(\ref{Equation_28}), for the special case $W_L=W_R=0$ and at $T=0$. The values of numerical solutions of the chemical potentials and the exact Fermi levels in the non-interacting BLGs are shown in the picture.}.
	\end{center}
\end{figure}

%
The time dependence of the total excitonic Josephson current through the BLG/I/BLG heterostructure, given in Eq.(\ref{Equation_28}), above, is evaluated numerically in Fig.~\ref{fig:Fig_6}, for different values of the applied gate potential, interaction parameters and condensates phase difference $\Delta\varphi_{0}=\varphi_{L}-\varphi_{R}$. When calculating numerically the principal value in Eq.(\ref{Equation_40}),  three singular points $(-1;0;1)$ of the integrand have been straddled properly. For the left-BLG, we have $W_L=0.5\gamma_0$ in all panels in Fig.~\ref{fig:Fig_6}, while for the right-BLG we have chosen three different values $W_R=0; 0.5\gamma_0$ and $W_R=\gamma_0$, from top to bottom panels in Fig.~\ref{fig:Fig_6}.
%
At $V=0$ and $\Delta{\varphi}_{0}=0$, no tunneling current flows in the system (see the solid black lines on the time axis with the holly triangular plot-markers), while a dc current appears at $V=0$ if the phases of coherent excitonic states, in different sides of the junction, differ by $\Delta\varphi_{0}=\pi/2$ or $\Delta\varphi_{0}=-\pi/2$ (see the solid black lines in Fig.~\ref{fig:Fig_6}). Remarkably, the dc excitonic Josephson current changes the sign when changing the sign of $\Delta\varphi_{0}$, i.e., $\Delta\varphi_{0}\rightarrow -\Delta\varphi_{0} \Rightarrow I_{\rm Exc}(t)\rightarrow -I_{\rm Exc}(t)$. Such a finite dc current in the system suggests the degeneracy in the ground state of the system, i.e., the U(1) symmetry, and the presence of the excitonic condensates. 
%
 \begin{figure}
 	\begin{center}
 		\includegraphics[width=240px,height=220px]{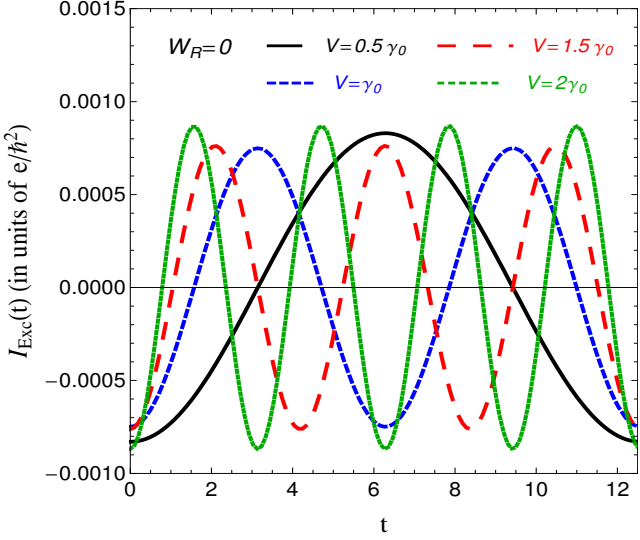}
 		\caption{\label{fig:Fig_7}(Color online) The time dependence of the total excitonic Josephson tunnelling current, given in Eq.(\ref{Equation_28}), for $W_R=0$ and for different values of the applied gate potential $V$. The condensates phase difference is fixed at the value ${\Delta\varphi}_{0}=\pi/2$ and the zero temperature case is considered here.}.
 	\end{center}
 \end{figure} 
 %
 Moreover, an ac excitonic current appears in the junction for any finite value of $V$ ($V=0.5\gamma_0$ and $V=\gamma_0$, in the picture), even for the case $\Delta\varphi_{0}=0$ (see the blue dotted lines in Fig.~\ref{fig:Fig_6} with the square plot-markers). The additional phase difference $\Delta{\varphi}_{0}=\pi/2$ only amplifies the excitonic tunneling current amplitude and leads to a phase shift (see the black dashed lines in Fig.~\ref{fig:Fig_6}). We observe also that the amplitude of the tunnel current decreases when increasing the parameter $W_R$ (see the top panel with $W_R=0$ and the middle panel with $W_R=0.5\gamma_0$, in Fig.~\ref{fig:Fig_6}). In the bottom pannel in Fig.~\ref{fig:Fig_6} we have chosen larger value of the applied gate potential $V=1.5\gamma_0$, and $W_R=\gamma_0$. We see that the oscillations of the tunnel current are multiplied in this case within the same time interval. This effect is clearly seen in Fig.~\ref{fig:Fig_7}, where different values of the applied gate voltage are considered straightforwardly, and the interaction parameter $W_R$ is fixed at the value $W_R=0$, in correspondence with the plots of the excitonic Josephson current in the upper panel in Fig.~\ref{fig:Fig_6}. We see that when multiplying the value of $V$ by an integer number $V'=nV$, where $n=1,2,3,4$, we have for the current wavelength $\lambda'=\lambda/n$, thus a relation of type $\lambda{V}=\text{const}$, between the applied gate potential and the current wavelength, emerges naturally.
  In Fig.~\ref{fig:Fig_8}, the same function $I_{\rm Exc}(t)$ is presented for the case of the fixed applied gate potential $V=0.5\gamma_0$, and for different values of the right-BLG interlayer Coulomb interaction parameter $W_R$, below the critical CNP value $W^{C}_{R}=1.48999\gamma_0$. The condensates phase difference is fixed at the value $\Delta{\varphi}_{0}=\pi/2$. We see, particularly, that when augmenting the interaction parameter $W_R$ (and keeping at the same time $W_L$ fixed at $W_L=0.5\gamma_0$) the amplitude of the excitonic tunnelling current is decreasing considerably. The zeros of the current function do not shift in their positions, contrary to the case presented in Fig.~\ref{fig:Fig_7}, where the shift of the zeros is caused by the reduction of the current wavelength in the junction. Thus, we realize that the interlayer Coulomb interaction in the right-BLG affects only the current amplitudes, while the changes of the applied gate potential modify principally the frequency of the excitonic current and have not a significant effect on the current amplitudes.   
%
%
 Next, in Fig.~\ref{fig:Fig_9}, we have shown the time dependence of the excitonic tunneling current for the values of $W_R$ above the UB charge neutrality point $W^{C}_{R}(UB)$. Contrary to the case, given in  Fig.~\ref{fig:Fig_8}, the amplitude of the Josephson tunneling current is increasing with $W_R$. This result is related again to the behavior of the chemical potential and Fermi energies in the BLGs (see in Ref.\onlinecite{cite_38}). In order to compare the results in Figs.~\ref{fig:Fig_8} and  ~\ref{fig:Fig_9}, we kept the curve for $W_R=0$, in both cases. 
 %
%
In Fig.~\ref{fig:Fig_10}, we have presented the excitonic Josephson current-voltage dependence on time at $T=0$ and for the balanced values of the interaction parameters $W_L$ and $W_R$. We put $W_L=W_R=0.5\gamma_0$ and we have considered the cases when $\Delta{\varphi}_{0}=0$ or $\Delta{\varphi}_{0}\neq 0$. We observe, in the upper panel in  Fig.~\ref{fig:Fig_10}, that for the case $\Delta{\varphi}_{0}=\pi/2$ we have a finite excitonic Josephson tunneling current through the system at $V=0$. At $V\neq 0$, we have principally the same qualitative behavior of the tunneling current for both cases $\Delta{\varphi}_{0}=0$ and $\Delta{\varphi}_{0}=\pi/2$ apart the situation at $t=0$, when there is a large threshold of $V$ at the zero phase difference (see the solid black curve in the lower panel, in Fig.~\ref{fig:Fig_10}). Contrary, for $\Delta{\varphi}_{0}=\pi/2$ and at $t=0$ the excitonic current has been developed in the system (see the solid black curve, in the upper panel, in Fig.~\ref{fig:Fig_10}). 
%
%
The $I-V$ characteristics of the excitonic Josephson current, for the case of the equal values of the interlayer interaction parameters, i.e., $W_L=W_R=W$, is shown in  Fig.~\ref{fig:Fig_11}. Different values of $W$ are considered in the picture: $W=0.5\gamma_0$ (solid black curve), $W=0.8\gamma_0$ (solid blue curve), $W=\gamma_0$ (solid yellow curve), $W=W^{\ast}=1.25\gamma_0$ (solid red curve), $W=1.3\gamma_0$ (dot-dashed darker green curve), $W=1.4\gamma_0$ (large-dashed green curve), and $W=W^{C}(LB)=1.48999\gamma_0$ (bold-dashed darker yellow curve). We observe that up to the value $W^{\ast}=1.25\gamma_0$, which corresponds to the maximum of the excitonic gap parameters in the BLGs, the amplitude of the excitonic Josephson current is increasing (including the dc values at $V=0$). Furthermore, for $W>W^{\ast}$, the amplitudes are continuously decreasing, for $W$ up to the LB CNP value $W=W^{C}(LB)=1.48999\gamma_0$. The further increase of $W$, above the LB CNP, leads to a drastic decrease (of about of one order of magnitude) of the excitonic tunnel current amplitude and this is shown in Fig.~\ref{fig:Fig_12}. It is important to concentrate on another principal difference between the results presented in Figs.~\ref{fig:Fig_11} and ~\ref{fig:Fig_12}. This concerns the first deepest minima of the excitonic Josephson current. In Fig.~\ref{fig:Fig_11}, those minima appear for relatively small voltages, and the positions of the positive (negative) minima are shifting to right (left), for $W\leq W^{\ast}=1.25\gamma_0$. Then, with further increase of $W$ in the interval $W^{\ast}<W<W^{C}(LB)=1.48999\gamma_0$, they are slightly shifting to left (right). On the other hand, the first minima for the curves in Fig.~\ref{fig:Fig_12}, for $W^{C}(UB)\leq W\leq3\gamma_0$, appear for very large values of $V$, and the positions of positive (negative) minima are continuously shifting to right (left) in this case. Therefore, the very large values of $V$ are very promising to observe the ac excitonic Josephson current in the system. Experimentally, this could be achieved with the appropriate choice of the left and right gate voltages, which will change the interlayer Coulomb interactions in the BLGs, until the expected effect takes place. In Figs.~\ref{fig:Fig_13}, we have presented the excitonic tunneling current for the asymmetric values of the interaction parameters $W_L$ and $W_R$ and we see how the increase of $W_R$ in the right-BLG, above the symmetric value $W_{R}=0.5\gamma_0$ (the parameter $W_L$ is fixed at the value $W_L=0.5\gamma_0$ for all values of $W_{R}$), leads to the tunneling current transfer into right, on the positive axis of $V$. It is interesting and straightforward to consider the special case of the non-interacting BLGs and the excitonic Josephson current through the junction in this particular case. This result is shown in Figs.~\ref{fig:Fig_14}. We see that although the zero interaction limit, the excitonic current still present in the BLG/I/BLG junction, and the values of it are comparable to the case when $W>W^{C}(UB)$, shown in Fig.~\ref{fig:Fig_12}, above. The reason of such an excitonic effect at $W=0$ is related to the finite solution of the chemical potential and the Fermi energy in the separate BLGs (see in Ref.\onlinecite{cite_38}, for more details).            
%
%
%
%
%
\section{\label{sec:Section_5} Concluding remarks}
%
We have calculated the normal quaiparticle and excitonic Josephson tunneling currents in the BLG/I/BLG heterostructure with the half-filled, AB-stacked bilayer graphene structures at different sides of the junction. By using the $S$-matrix approximation, we have derived the analytical expressions of both types of tunneling currents: normal quasiparticle and excitonic. The role of charge neutrality point has been discussed in details. Particularly, we have shown how the current spectrum is changing when passing through the CNP point of the interlayer Coulomb interaction parameters in different sides of the junction. It has been shown that the normal quasiparticle tunneling in the junction is a threshold process, independently of the values of the interlayer Coulomb interactions in the BLGs and the field frequency threshold could be modified by changing the interaction parameter in one BLG (for example the right), while keeping it fixed, at the same time, in another BLG in the junction. 
The formation of the resonant tunneling peaks in the normal tunneling spectrum has been analyzed in details by considering the vicinity of the charge neutrality point in the right-BLG. The very large values of the threshold frequency have been obtained for the case of the interaction-symmetric junction $W_L=W_R$, below and above the charge neutrality point. 
Particularly, from the form of $I-V$ spectrum, in the case of the interaction-asymmetric junction, it is clear that the excitonic insulator state and the excitonic condensate states are two different states of matter in the BLGs. We have shown that the low-frequency peaks  appear from the excitonic tunneling in the condensates regime in BLGs, and at the large values of the Coulomb interaction parameter in the right-BLG (when the excitonic insulator state breaks down). The temperature dependence of the field frequency threshold values has been derived for a very large interval of temperature: from zero up to very high temperatures. 

Furthermore, the time dependence of the excitonic Josephson tunnel current has been studied in details for different limits of interlayer Coulomb interaction parameter in the right-BLG, and the excitonic dc effect has been pointed out for the nonzero phase difference $\Delta{\varphi}_{0}\neq 0$ and $V=0$. Ulteriorly, an ac excitonic Josephson effect has been shown in the junction for non-zero values of the applied gate potential. It has been shown that the phase difference between the excitonic condensates in the BLGs, only amplifies the tunneling current when keeping $V$ constant . It has been shown that a relation of type $\lambda{V}=\text{const}$ is valid in the junction, when changing the applied gate potential by the integer values. The time dependence of the excitonic Josephson tunneling current has been analysed by considering different values of the right-BLG interaction parameter and the role of the CNP point has been revealed out. The $I-V$ characteristics of the excitonic Josephson effect has been found for both interaction-symmetric and interaction-asymmetric cases, and the effects of the charge neutrality point have been discussed in details. It has been shown that the very large values of $V$ are very promising to observe the ac excitonic Josephson current in the system. Finally, a particular case of the non-interacting BLGs junction has been considered separately, and the excitonic Josephson tunneling current has been calculated for this case. As a result, the non zero excitonic current is shown present in this case with the amplitudes comparable to that of the amplitudes of $I-V$ spectrum at the high symmetric values of $W$ in different sides of the junction. 

We have presented a self-exhausting theory of the bilayer graphene based heterojunction and the results presented here could represent a veritable framework on which the experimental setup could be made and the results could be compared. The theoretical results in the paper are especially important in the context of the long standing problem about the excitonic condensation and pair formation in the BLGs and double layer graphene systems. This is especially important for theoretical understanding of the nature of the EI state and the coherent excitonic condensate states in such systems: their similarities and differences. The Josephson effect studied in the considered system could be furthermore an important construction in order to build the graphene-based quantum interference devices, which will bring a new idea about the ultra-sensitive magnetometers and voltmeters and will be more sensitive than the usual superconductors-based SQUIDs due to the exceptional mobility of the electrons in graphene. The excitonic Josephson junction is also promising in the context of building a new type of ultrafast electronic circuits blocks, which will form a digital logic unit in the modern ultrafast computers and fast electronics.      
\appendix


%
\end{document}